\documentclass[journal]{IEEEtran}
\usepackage{pgfplots}
\pgfplotsset{compat=1.18}
\colorlet{Mcol}{black!55}
\colorlet{shellcol}{cyan!12}
\colorlet{shelledge}{cyan!55!black}
\colorlet{dcol}{blue!65!black}
\colorlet{acol}{red!70!black}
\colorlet{okcol}{green!55!black}
\tikzset{
      vec/.style={-{Latex[length=2mm,width=1.4mm]}, line width=0.7pt},
      thinvec/.style={-{Latex[length=1.6mm,width=1.1mm]}, line width=0.5pt},
      pt/.style={circle,fill=black,inner sep=1.05pt},
      opt/.style={circle,draw=black,fill=white,inner sep=1.0pt,line width=0.5pt},
      lbl/.style={font=\scriptsize,inner sep=6pt},
    }

\usepackage{comment}
\usepackage[backend=biber,style=ieee,maxbibnames=6]{biblatex}
\usepackage{amsmath,amssymb,amsfonts}
\PassOptionsToPackage{dvipsnames,svgnames,table}{xcolor}
\usepackage{multirow}
\usepackage{graphicx}
\usepackage{textcomp}
\usepackage{xcolor}
\usepackage{array}
\usepackage{booktabs}
\usepackage{subcaption}
\usepackage{placeins}        
\usepackage{stfloats}        

\usepackage{algorithm}
\usepackage{algpseudocode}
\usepackage{hyperref}

\usepackage{tikz}
\usetikzlibrary{arrows.meta,decorations.pathreplacing,calc,positioning}
\usepackage{tabularx}
\usepackage{amsthm}
\theoremstyle{plain}
\newtheorem{proposition}{Proposition}

\theoremstyle{remark}
\newtheorem{remark}{Remark}
\newtheorem{lemma}{Lemma}
\newtheorem{theorem}{Theorem}
\newtheorem{assumption}{Assumption}
\newtheorem{fact}{Fact}
\newtheorem{corollary}{Corollary}

\usepackage{capt-of}

\newcolumntype{P}[1]{>{\centering\arraybackslash}p{#1}}
\def\BibTeX{{\rm B\kern-.05em{\sc i\kern-.025em b}\kern-.08em
    T\kern-.1667em\lower.7ex\hbox{E}\kern-.125emX}}

\begin{document}
\title{Model-Free Generative Stealthy Attacks on Safety-Critical Vehicle Cyber-Physical Systems}








\author{Qazi~Mairaj~ud~din,
Sidra~Ghayour~Bhatti,
and~Qadeer~Ahmed%
\thanks{The authors are with the Center for Automotive Research, The Ohio State University, Columbus, OH, USA (e-mail: mairajuddin.1@osu.edu;
bhatti.39@osu.edu; ahmed.358@osu.edu). Implementation details, trained models, and evaluation scripts available at: \url{https://github.com/OSU-CAR-MSL/model_free_stealthy_attack}.}}

\maketitle

\begin{abstract}

Existing stealthy attacks presuppose white- or grey-box access to the target system, unrealistic in practice. This paper develops a model-free stealthy attack framework that learns an attack policy directly from nominal measurement data, without knowledge of the system dynamics, estimator, or detector. We establish a certified residual--impact bound and an observable stealth budget about a stealth-consistent anchor, and show recurrence and context-conditioning are structural necessities for closed-loop, multivariate Cyber-Physical Systems (CPS). The resulting attack generator, trained against a Wasserstein critic under a stealth-constrained impact objective, induces a hazardous 2.51\,m lane departure while evading five intrusion detection systems, and, in real-time hardware-in-the-loop tests, an $\approx\!8$\,km/h speed deviation, while staying below the detection threshold. Results show safety-critical CPS remain exposed without attacker access to privileged information, motivating offense-informed defenses.


\end{abstract}



\section{Introduction}


Cyber-Physical Systems are increasingly evolving, couple sensing, computation, communication, and control with physical processes, driven by Industry 4.0 and growing connectivity across industrial control, smart grid, medical device, and vehicle systems. This growth has introduced new security
challenges for safety-critical operation, illustrated by incidents from Stuxnet~\cite{langner2011stuxnet} and the Ukrainian power grid~\cite{liang20162015} to the remote Jeep
hack~\cite{miller2015remote} and, in automotive networks specifically, CAN-injection vehicle
theft~\cite{Tindell2023CANInjection}, fleet cyber-worms~\cite{jepson2024commercial}, and the 2026 Intoxalock
outage affecting $\sim$150{,}000 interlock-equipped vehicles~\cite{wired2026intoxalock}. These indicate CPS
attack surfaces are outpacing defenses, motivating analysis that generalizes across CPS domains.

The security of CPSs has been widely studied from the dual perspectives of the defender and the adversary, because a precise understanding of worst-case attack behavior is a prerequisite for principled defense design \cite{guo2022stealthy, jin2025completely}. Within this body of work, this work concerns the adversarial perspective. 

Early theoretical work established the feasibility of \emph{stealthy} attacks in CPS under idealized white-box conditions. Mo and Sinopoli~\cite{mo2010fdia,mo2015performance} showed that an adversary with full knowledge of the plant, estimator, observer gains, and noise statistics can construct attacks that preserve nominal innovation-residual behavior and
evade detection. These frameworks derive necessary and sufficient conditions for perfect attackability, but are formulated for abstract Linear Time-Invariant (LTI) systems under simplified plant, sensing, and communication assumptions, with empirical validation typically limited to abstract numerical examples that do not capture real-world CPS complexity. A second line of work addressed this abstraction gap with stealthy attack frameworks tailored to application domains---smart grids~\cite{hug2012vulnerability,
hao2015sparse, guo2016optimal}, industrial control systems~\cite{jiang2023monitoring,
ma2019stealthy, urbina2016limiting}, and vehicle platoons~\cite{ahmed2024detection,
ju2020deception}. While these demonstrate practical relevance, they continue to rely on white-box assumptions---precise plant model, estimator gains, noise covariances, and detector thresholds---that in practice demand significant
adversary resources to acquire. ISO/SAE 21434~\cite{iso21434_2021} formalizes
this intuition, characterizing \emph{attack feasibility} via attacker \emph{(i) expertise}, \emph{(ii) system knowledge}, and \emph{(iii) window of opportunity}; white-box frameworks implicitly assume the adversary scores highly on all three. Recent research has shifted toward grey-box and data-driven attack generation, reducing privileged-knowledge requirements. Khazraei et al.~\cite{khazraei2022learning}
proposed learning-enabled attack generators using feed-forward (FNN) and recurrent (RNN) neural networks in a grey-box setting, degrading CPS performance. Parameter- and structure-agnostic techniques
followed~\cite{costilla2022attack, xiao2025guaranteed, jiao2021new}
, alongside deep-learning based attack generation~\cite{nawaz2025deep, narang2024physical} and multi-sensor remote-estimation attacks exploiting side information or partial observability~\cite{li2024optimal,
guo2025stealthy}, marking a clear trend from white-box toward data-driven attack models. Our own prior work \emph{GAttack}~\cite{mairaj2024gattack} is also aligned with the data-driven shift where we introduced an LSTM-driven GAN synthesized temporally coherent in-vehicle payloads that bypassed a state-of-the-art IDS~\cite{kukkala2020indra}.

Despite this progression, existing stealthy-attack frameworks transfer poorly to heterogeneous, dynamic CPS:
(i) they remain \emph{white- or grey-box}, presupposing privileged, operator-confidential information;
(ii) they are validated on low-order, linear, or univariate benchmarks, ignoring the joint distribution and
temporal dependence of correlated multivariate measurements; (iii) their generators assume prolonged offline
access to closed-loop and detector-relevant signals; and (iv) evaluation is confined to quasi-static,
centralized, smart-grid regimes under narrowly varying conditions, with no robustness guarantee across dynamic operation.

\begin{figure}[t]
    \centering
    \includegraphics[width=1.05
    \linewidth]{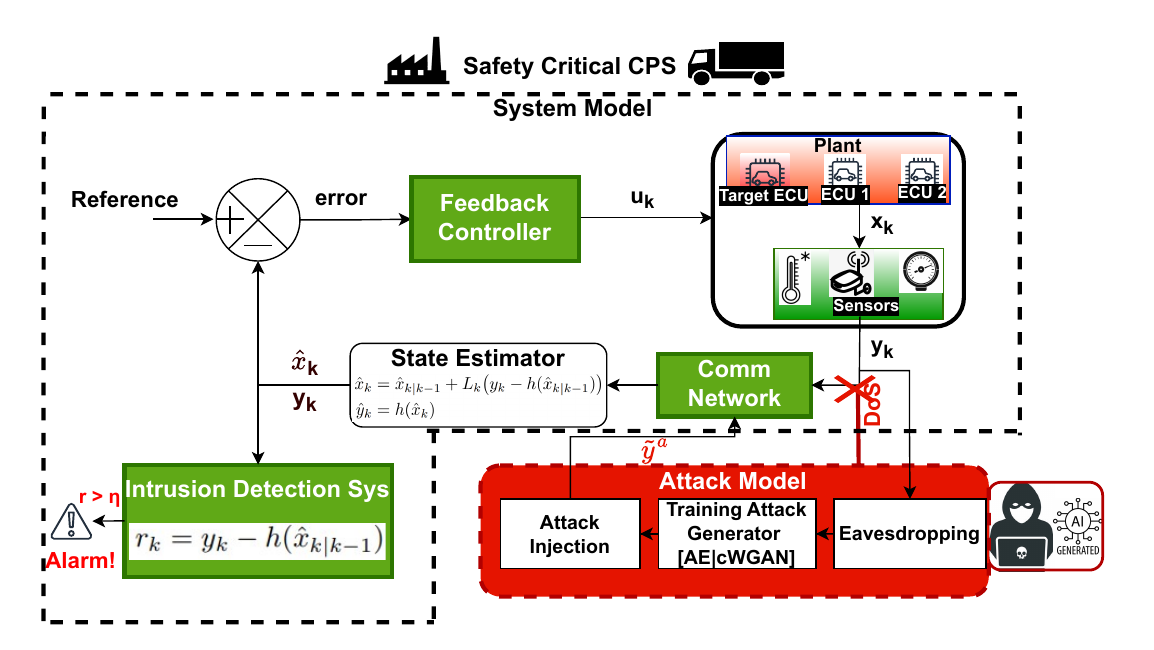}
    \caption{Typical CPS architecture and proposed Stealthy Attack Model.}
    \label{fig:cgattack}
\end{figure}

These gaps motivate a model-free stealthy attack formulation. Fig.~\ref{fig:cgattack} shows a typical CPS architecture and our proposed attack model. The main contributions are:

\begin{enumerate}

\item We \emph{formalize} residual-based detectability as a property of the observer-induced measurement \emph{manifold}, extending the classical linear column-space undetectability condition to the nonlinear recursive closed-loop setting. This grounds the attack in an object recoverable from data rather than explicit system knowledge, which a model-free adversary never observes.

\item We \emph{derive} a certified bound that lower-bounds the unobservable state-estimation error by the attacker-observable injection magnitude for arbitrary Lipschitz sensing maps without plant, estimator, or detector model. We further \emph{prove} that a per-sample residual statistically indistinguishable from nominal is integrated by the closed loop into an unbounded accumulated deviation, establishing a regime in which stealth and impact are simultaneously attainable.

 \item We \emph{establish} an observable certified stealth budget about a stealth-consistent anchor via a Rayleigh-quotient bound, and \emph{realize} the anchor as a regularized autoencoder whose reconstruction residual estimates the manifold distance of the nominal density, with a bottleneck matched to the sensor-set observability limit (Hermann--Krener rank). This replaces the model-dependent innovation covariance $S_k$ with an empirical nominal quantile, eliminating any Jacobian or detector access.

\item We \emph{prove} that recurrence and context-channel conditioning are structural necessities for holding the stealth budget while inducing impact in closed-loop, multivariate CPS, and \emph{synthesize} both into a recurrent, context-conditioned cWGAN-GP generator. Trained on nominal data alone, it \emph{generalizes} across platforms and control channels: a $2.51$\,m lane departure on a benchmark EV model and an $\approx 8$\,km/h speed deviation on a real-time dSPACE\,/\,J1939 truck-in-the-loop platform, while evading five state-of-the-art IDSs.

\end{enumerate}

The remainder of this paper is organized as follows. Section~\ref{sec:system_attack} formalizes the system and attack model. Section~\ref{sec:methodology} presents the proposed framework
. Section~\ref{sec:experiments} reports experimental results and ablation studies. Section~\ref{sec:conclusion} concludes.

\section{System Definition and Attack Model}\label{sec:system_attack}
This section formalizes general nonlinear CPS architecture and attack model shown in Fig.~\ref{fig:cgattack}.

\subsection{System Model}\label{subsec:system_model}

We consider a discrete–time nonlinear dynamical system,
\begin{equation}
    x_{k+1} = f(x_k, u_k) + w_k,
    \label{eq:dynamics}
\end{equation}
where $x_k \in \mathbb{R}^n$ denotes the internal states, $u_k \in \mathbb{R}^m$ is the control input, and $w_k \sim \mathcal{N}(0,Q)$ is process noise. Sensor measurements available on the network are modeled as
\begin{equation}
    y_k = h(x_k) + v_k,
    \label{eq:output}
\end{equation}
where $y_k \in \mathbb{R}^p$ denotes plant output (sensor measurements); $v_k \sim \mathcal{N}(0,R)$ is measurement noise. The mapping $h(\cdot)$ represents the nonlinear sensor model and is assumed to be Lipschitz continuous, consistent with standard observer design assumptions \cite{661604}. Three standard CPS components are considered:

\paragraph*{State Estimator.}
A recursive estimator (KF~\cite{kalman1960new} for LTI, 
EKF for nonlinear systems) provides state estimates via the 
prediction--correction recursion:
\begin{equation}
\begin{aligned}
\hat{x}_{k|k-1} &= f(\hat{x}_{k-1}, u_{k-1}), \\
\hat{x}_{k} &= \hat{x}_{k|k-1} + L_k 
              \underbrace{\big(y_k - h(\hat{x}_{k|k-1})\big)}_{r_k},
\quad \hat{y}_{k} = h(\hat{x}_{k}),
\end{aligned}
\label{eq:kalman_predict}
\end{equation}
where $L_k$ is the filter gain. The innovation residual,
\begin{equation}
    r_k = y_k - h(\hat{x}_{k|k-1}),
    \label{eq:residual}
\end{equation}
is the key signal monitored by the IDS. Any sustained sensor 
corruption that biases $\hat{x}_k$ propagates directly into 
erroneous control actions and degrading closed-loop performance.

\paragraph*{Feedback Controller.} The control input is generated by a feedback policy $u_k=\pi(\hat{x}_k)$, where $\pi$ is the control action depending on the state estimate.

\paragraph*{Intrusion Detection System.}
Under nominal Gaussian noise, $r_k$ is zero-mean with covariance 
$S_k = C_k P_{k|k-1} C_k^\top + R$, and a $\chi^2$-based detector 
\cite{mo2015performance, mo2010fdia} evaluates the quadratic 
innovation statistic,
\begin{equation}
    g_k = r_k^\top S_k^{-1} r_k,
    \label{eq:innovation_stat}
\end{equation}
triggering an alarm when $g_k > \eta$. Since real-world sensor 
data frequently exhibit non-Gaussian heavy tails that 
miscalibrate the $\chi^2$ threshold~\cite{Hwang2010FDIRSurvey}, 
we instead adopt a non-parametric, percentile-based detector whose 
threshold is computed from attack-free nominal data:
\begin{equation}
    \eta = \operatorname{Percentile}\!\left(
    \{ g_j \}_{j \in \mathcal{N}},\; 1-\epsilon
    \right),
    \label{eq:percentile_threshold}
\end{equation}
where $\mathcal{N}$ is a nominal window and $\epsilon$ is the 
target false-alarm rate. This detector generalizes the $\chi^2$ 
test (converging to it under Gaussianity) and can be applied 
consistently across both simulated and real-world datasets. A state estimator produces $\hat x_k$ from $\{y_j,u_j\}_{j\le k}$, and a
residual-based detector computes $r_k$~\eqref{eq:residual} and
$g_k$~\eqref{eq:innovation_stat}, alarming when $g_k>\eta$. The attack developed in this work targets this residual-energy test. We denote by $\Gamma_{\mathrm{mag}}$ the \emph{magnitude} (energy) detector class: any test that scores the quadratic energy of the innovation, normalized by the model covariance. This class includes (i) the per-step $\chi^2$/NIS test $g_k=r_k^\top S_k^{-1}r_k>\eta$ of~\eqref{eq:innovation_stat}; (ii) its
windowed exceedance-count variant, which alarms on the number of threshold crossings in a sliding window; and (iii) CUSUM/energy tests on $\|r_k\|$. Stealth throughout this paper is established against $\Gamma_{\mathrm{mag}}$: the attack holds the normalized innovation energy at its nominal operating point~\eqref{eq:percentile_threshold} while inducing impact.

\subsection{Attack Model}\label{subsec:attack_model}
In a realistic setting, the attack is \emph{resource-constrained}
\cite{guo2022stealthy}: only sensors in a compromised set
$\mathcal K_a\subset\{1,\dots,p\}$, with $|\mathcal K_a|=p_a$, can be manipulated. We partition the sensor index set as
$\{1,\dots,p\} = \mathcal K_a \,\dot\cup\, \mathcal K_a^c$, and the output map, measurement, and noise accordingly as $h = [h_a;h_\xi]$, $y_k = [y_k^a;\xi_k]$, $v_k=[v_k^a;v_k^\xi]$, giving the nominal (pre-attack) blocks
\begin{equation}
y_k^a = h_a(x_k) + v_k^a \in\mathbb R^{p_a},
\qquad
\xi_k = h_\xi(x_k) + v_k^\xi \in\mathbb R^{p-p_a},
\label{eq:partition}
\end{equation}
the target (compromised) channel $y_k^a$ carrying the injected attack and the context channel $\xi_k$ the remaining uncompromised measurements. The adversary injects an additive vector $a_k\in\mathbb R^{p_a}$ on the target block, $\tilde y_k^a := y_k^a + a_k$, leaving $\xi_k$ unmodified, so the estimator/controller receive $\tilde y_k = [\tilde y_k^a;\xi_k]$.

The adversary's capabilities are consolidated into realistic assumptions for legacy and remotely accessible CPS networks: (i) it can read and overwrite frames due to no message authentication and integrity protection~\cite{jepson2024commercial,Tindell2023CANInjection},
and decode them into physical signals via public specifications (e.g.\ SAE~J1939) or reverse-engineered databases (e.g.\ opendbc); (ii) it can passively eavesdrop long, high-quality recordings of attack-free operation while undetected~\cite{7467419}; and
(iii) the operating distribution at deployment matches that of the nominal recording. Correspondingly, the attack requires \emph{no} plant model $f(\cdot)$, sensing map $h(\cdot)$, Jacobians, estimator internals, or detector parameters and all quantities
are recovered from the nominal recordings.

Formulating the problem, the adversary seeks to maximize the induced state-estimation error while remaining below the detector's threshold. We define a twofold objective:
\begin{equation}
\max_{\text{}}\;\;\underbrace{\mathbb E\big\{\|\Delta x_k\|\big\}}_{\text{Impact}}
\;
\qquad\text{s.t.}\qquad
\underbrace{g_k \le \eta}_{\text{Stealth}} \ .
\label{eq:abstract_problem}
\end{equation}

\begin{enumerate}
    \item \textbf{Impact:} inducing state-estimation error \emph{$\Delta x_k$} to degrade closed-loop performance, which propagates into incorrect control actions $u_k=\pi(\hat{x}_k)$ and perturbs the state trajectory.
    \item \textbf{Stealth:} avoid triggering the residual-based detector that alarms when the innovation statistic $g_k$ exceeds a threshold $\eta$.
\end{enumerate}

\section{Proposed Methodology}\label{sec:methodology}

\subsection{Model-Free Attack Framework}
\label{sec:overview}

Under the \emph{model-free} constraint of Section~\ref{subsec:attack_model}, the attacker
observes \emph{neither} quantity in~\eqref{eq:abstract_problem} since they depend on the plant map $h$, dynamics $f$, and estimator gains. Both the objective and the constraint must therefore be re-expressed in quantities observable from nominal data alone. We approach this by establishing what \emph{exact} stealth demands of a fully model-aware adversary which is settled in the literature, and arrive at a regime 
in which stealth and impact are simultaneously attainable.

\begin{lemma}[Static residual invariance; Liu \emph{et al.}~{\cite[Thm.~3.1]{liu2011false}}]
\label{lem:static}
Consider the static measurement model $y = Hx + v$ with a memoryless (weighted least-squares) estimate $\hat{x} = (H^\top W H)^{-1} H^\top W y$ and residual $r = y - H\hat{x}$.
For any $c\in\mathbb{R}^{n}$, the attack $a = Hc$ leaves $\|r\|$ exactly unchanged and
shifts the estimate by exactly $c$. Detectability is thus governed by the column space
$\mathcal{M} = \mathrm{col}(H)$, which is invariant to any right factorization $H'=HT$ with $T$
invertible.
\end{lemma}

Lemma~\ref{lem:static} identifies the geometric object that governs stealth, but it is stated for a \emph{memoryless} estimator. However, the detector in~\ref{sec:system_attack}, is not memoryless and the reference $h(\hat{x}_{k|k-1})$ is propagated through the plant dynamics, so an offset that is admissible at one instant need not remain admissible at the next. Hence, exact stealth against a recursive residual test imposes two conditions. i) the injection must lie in the measurement range at each instant, \emph{and} ii) the underlying offset $c_k$ must itself evolve as $c_{k+1} = A c_k$. In the nonlinear setting~\eqref{eq:dynamics}--\eqref{eq:output} the reachable measurement set $\mathcal M$ is a curved manifold rather than a fixed subspace; the superposition underlying both the range condition of Lemma~\ref{lem:static} and the linear offset recursion $c_{k+1}=Ac_k$ no longer holds, and no attack in general reproduces a state-shifted measurement exactly once $h$ is nonlinear. Locally, the first-order approximation $h(x)\approx H_kx+b_k$ with $H_k=\nabla_x h|_{x_k}$ recovers the linear structure in a neighborhood of each operating point, motivating manifold-aligned attack design.

The remainder of this section derives four requirements for a model-free attack, and realizes each with a trainable component:
\begin{itemize}
\item[(R1)] a stealth-consistent \emph{anchor} against which the detector's residual is assessed, yielding an observable stealth \emph{budget} on the displacement about it where a tolerance must be measured against a reference the adversary can compute;
\item[(R2)] an \emph{observable impact proxy} that provably lower-bounds the true, unobservable impact, so that~\eqref{eq:abstract_problem} can be optimized without the
plant model and maximized within the R1 budget;
\item[(R3)] a \emph{recurrent} attack law: by
Theorem~\ref{thm:accumulation} the map from budget to impact is a plant-driven recursion, so a model-free adversary must learn a history-dependent law rather than a pointwise map;
\item[(R4)] \emph{conditioning on the uncompromised context channels}: partial compromise pins the context block of the residual deviation, so the perturbation  must be shaped against the joint nominal statistics.
\end{itemize}

\subsection{Framework Realization}
\label{sec:geometry_objective}

\begin{figure}[t]
    \centering
    \includegraphics[width=1.3\linewidth]{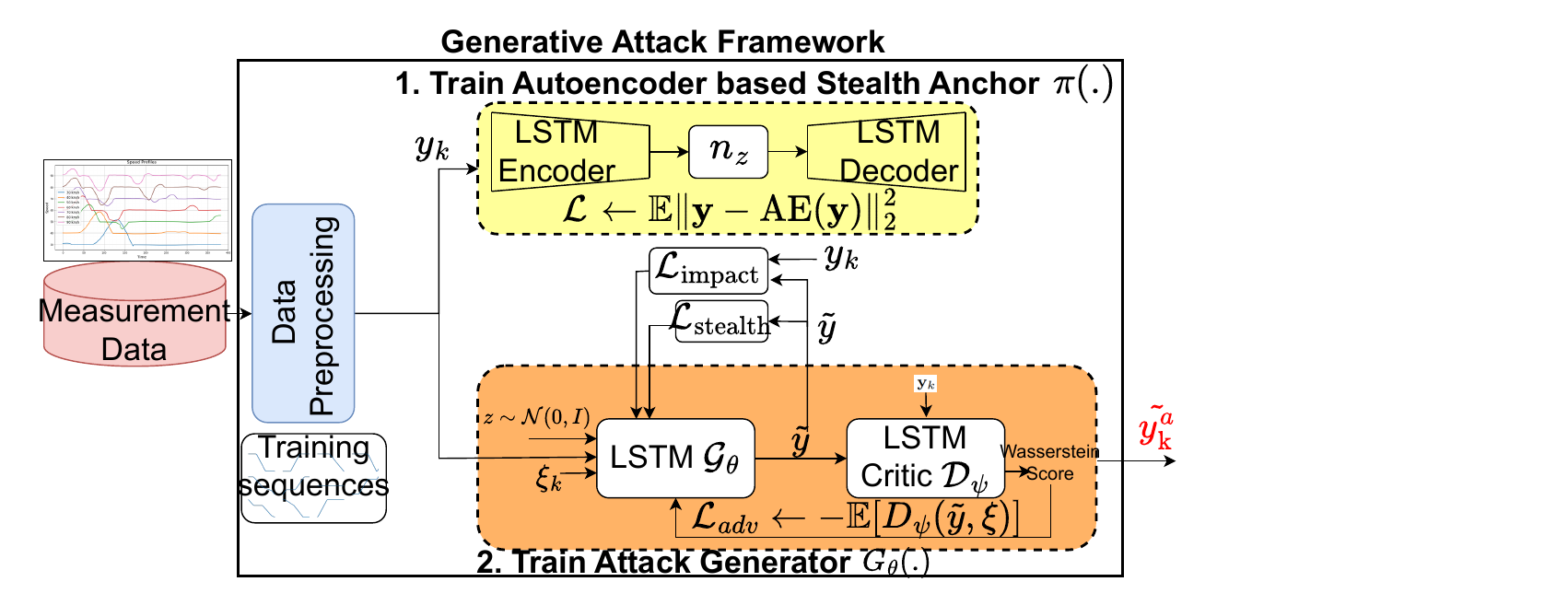}
    \caption{Overall architecture of the proposed attack framework.
    }
    \label{fig:overall_arch}
\end{figure}

We begin by making the objective in~\eqref{eq:abstract_problem} observable. Residual-based detectors assess whether incoming measurements are consistent with the observer-induced measurement manifold, rather than verifying the exact physical parameters of the plant (Lemma~\ref{lem:static}). Extended to the nonlinear
closed-loop setting, this yields a certified \emph{observable} impact proxy (Lemma~\ref{lem:residual_impact_coupling}) and an observable stealth budget about a stealth-consistent anchor (Proposition~\ref{prop:reflection_direction}). An autoencoder (AE) then instantiates that anchor from nominal data alone. The generator
$G_\theta$ observes the strictly-past target history and the available context, and emits a bounded \emph{displacement} about a frozen anchor $\Pi$ that enforces the stealth budget. The injected attack is the resulting deviation from the true measurement. Fig.~\ref{fig:overall_arch} shows the overall architecture.

The objective~\eqref{eq:abstract_problem} is stated on the unobservable $\Delta x_k$. The following result lets the attacker act on the observable residual instead, certifying that a large residual forces a large true error.

\begin{lemma}[Residual--impact coupling: proxy validity (R2)]
\label{lem:residual_impact_coupling}
Let $\Delta x_k := x_k - \hat x_{k|k-1}$ denote the (attacker-unobservable) prior
estimation error. Then, under the following assumption,

\begin{assumption}[Lipschitz measurement map, bounded noise]
\label{as:lipschitz}
The measurement map $h:\mathbb{R}^n\to\mathbb{R}^p$ is $L$-Lipschitz with constant $L$,
i.e.\ $\|h(x)-h(y)\|\le L\|x-y\|$ for all $x,y\in\mathbb{R}^n$, and the measurement
noise covariance satisfies $R \preceq \sigma I$ for some $\sigma > 0$, where $\sigma$ is
a scalar bound on the largest eigenvalue of the noise covariance matrix $R$, with $I$
the identity matrix.
\end{assumption}

for any residual level $\alpha>0$, measurement dimension $p$ ($y_k\in\mathbb R^p$), and
concentration parameter $\kappa>0$, the event
$\mathcal E_\kappa := \{\|v_k\|\le\sqrt{\sigma}\,\kappa\}$ has probability
$\mathbb P(\mathcal E_\kappa)\ge 1-p/\kappa^2$, and on this event
\begin{equation}
\|r_k\| \ge \alpha
\;\Longrightarrow\;
\|\Delta x_k\| \ge \frac{\alpha - \sqrt{\sigma}\,\kappa}{L}.
\label{eq:residual_impact_bound}
\end{equation}
The bound is non-vacuous when $\alpha > \sqrt{\sigma}\,\kappa$, i.e.\ when the residual
level exceeds the noise floor.
\end{lemma}

\begin{proof}
By the multivariate Chebyshev inequality~\cite{navarro2016very} applied to the
zero-mean measurement noise $v_k\in\mathbb R^p$,
$\mathbb{P}\big(v_k^\top R^{-1} v_k \le \kappa^2\big) \ge 1 - p/\kappa^2$. Since
$R \preceq \sigma I$, it holds that $\sigma^{-1} v_k^\top v_k \le v_k^\top R^{-1} v_k$
for all $v_k$, so the event $\mathcal E_\kappa := \{\|v_k\| \le \sqrt{\sigma}\,\kappa\}$
satisfies $\mathbb P(\mathcal E_\kappa)\ge 1-p/\kappa^2$. Conditioning on this event,
the triangle inequality gives
\begin{equation*}
\alpha \le \|r_k\| = \|h(x_k) + v_k - h(\hat x_{k|k-1})\|
\le L\|\Delta x_k\| + \sqrt{\sigma}\,\kappa,
\end{equation*}
using the $L$-Lipschitz property of $h$. Rearranging gives
$\|\Delta x_k\| \ge (\alpha - \sqrt{\sigma}\kappa)/L$ on $\mathcal E_\kappa$.
\end{proof}

\noindent\emph{Role of Lemma~\ref{lem:residual_impact_coupling}.} The bound certifies the
\emph{observable} residual as a valid proxy for the unobservable error: any injection
that raises $\|r_k\|$ above the noise floor raises a certified floor on $\|\Delta x_k\|$,
so maximizing the observable surrogate
$\mathcal L_{\mathrm{impact}}=\mathbb E\{\|\tilde y_k^a-y_k^a\|^2\}$
(Section~\ref{sec:lagrangian_objective}) optimizes a lower bound on the true impact.
By construction, however, the stealth threshold $\tau$ keeps $\|r_k\|$ near the noise
floor, so this per-sample floor is small; it establishes feasibility of the proxy, not the magnitude of the achievable impact.


\begin{theorem}[Integrated-bias impact]
\label{thm:accumulation}
Consider the linear specialization of~\eqref{eq:dynamics}--\eqref{eq:output}
($f(x,u)=Ax+Bu$, $h(x)=Cx$) in closed loop with steady-state Kalman gain $K$, and decompose the prior estimation error under attack as
$\Delta x_k=\Delta x_k^{\mathrm{nom}}+\varepsilon_k$, where $\Delta x_k^{\mathrm{nom}}$ is the attack-free prior error---zero-mean and independent of the injection---and $\varepsilon_k$ the attack-induced component.
\begin{enumerate}
\item[(i)] \emph{(Accumulation.)} Under \emph{any} injection sequence $\{a_k\}$,
\begin{equation}
\varepsilon_{k+1} \;=\; A\,\varepsilon_k \;-\; AK\,\Delta r_k ,
\label{eq:driven}
\end{equation}
independently of the injection itself; for an attack initiated at $k_0$ from nominal operation ($\varepsilon_{k_0}=0$),
\begin{equation}
\varepsilon_k \;=\; -\sum_{j=k_0}^{k-1} A^{\,k-j}\,K\,\Delta r_j .
\label{eq:convolution}
\end{equation}
\item[(ii)] \emph{(Modal amplification.)} Let $\Delta r_k$ have a constant mean component $\bar\delta$, and let $\lambda$ be an eigenvalue of $A$ with right/left eigenvectors $\phi,\psi$ normalized $\psi^\top\phi=1$. The bias projected on this mode is
\begin{equation}
\psi^\top \mathbb E[\varepsilon_k]
= -\,(\psi^\top K\bar\delta)\sum_{m=1}^{k-k_0}\lambda^{\,m},
\label{eq:modal_projection}
\end{equation}
which, with $c:=\psi^\top K\bar\delta$, (a) converges to $-c\,\lambda/(1-\lambda)$ if $|\lambda|<1$; (b) grows linearly, $-c\,(k-k_0)$, if $\lambda=1$ is semisimple, and as a
polynomial of degree equal to the Jordan index of $\lambda$ if $\lambda=1$ is defective;
(c) diverges geometrically if $|\lambda|>1$.
\item[(iii)] \emph{(Impact per detectability.)} The detector mean shift is the horizon-independent constant $\|\bar\delta\|_{S^{-1}}^2$ (Corollary~\ref{cor:cov_preservation}). Hence on a mode with $|\lambda|=1$ and $c\ne0$
the ratio of accumulated impact to detector mean shift grows without bound in the attack horizon, at the fixed nominal false-alarm rate.
\end{enumerate}
\end{theorem}

\begin{proof}
Let $e_k := x_k - \hat x_k$ be the posterior estimation error and $\Delta e_k$ deviation between the attacked and attack-free runs; the process and measurement noise are common to both runs and cancel. Eliminating the injection between the posterior-error and innovation deviation recursions~\cite[eqs.~(28)--(29)]{mo2010fdia} gives $\Delta e_{k+1}=A\,\Delta e_k-K\,\Delta r_{k+1}$. Since the prior error propagates the posterior error through the plant, $\varepsilon_k=A\,\Delta e_{k-1}$; substituting yields
$\varepsilon_{k+1}=A\Delta e_k=A(A\Delta e_{k-1}-K\Delta r_k)=A\varepsilon_k-AK\Delta r_k$,
which is~\eqref{eq:driven}, and~\eqref{eq:convolution} is its variation-of-constants solution with $\varepsilon_{k_0}=0$.
For (ii), left-multiplying~\eqref{eq:convolution} by $\psi^\top$ with $\Delta r_j\to\bar\delta$ gives \eqref{eq:modal_projection}; $\sum_{m=1}^{M}\lambda^m=\lambda(1-\lambda^M)/(1-\lambda)$ for $\lambda\ne1$ and $M$ for semisimple $\lambda=1$, while a defective $\lambda=1$ of Jordan index $d$ yields $\Theta(M^{d})$ growth along its top generalized eigenvector. 
Part (iii) combines (ii) with the horizon-independent mean shift of Corollary~\ref{cor:cov_preservation}.
\end{proof}

\begin{corollary}[Covariance preservation]
\label{cor:cov_preservation}
Let $S_k$ denote the innovation covariance of Section~\ref{subsec:system_model}, fixed
by the Riccati recursion and hence independent of the measurement stream and of the attack, with steady-state value $S$. For an additive injection whose residual deviation $\Delta r_k$ is \emph{past-measurable} (a deterministic function of information available before the current noise $v_k$, as realized by the deterministic DC bias and by the recurrent generator of Section~\ref{sec:attack_law}), the monitored statistic
$g_k=\tilde r_k^\top S^{-1}\tilde r_k$ with $\tilde r_k=r_k^{\mathrm{nom}}+\Delta r_k$
satisfies, to first order in the stealth budget,
\begin{equation}
\operatorname{Cov}(\tilde r_k)=S+O\!\big(\|\Delta r_k\|^2\big),
\qquad
\mathbb E[g_k]=p+\|\Delta r_k\|_{S^{-1}}^2 .
\label{eq:mean_shift}
\end{equation}
Consequently a detector calibrated to false-alarm rate $\epsilon$ on nominal data retains that rate on the attacked stream while $\|\Delta r_k\|_{S^{-1}}$ stays below the nominal
innovation floor. For a deterministic injection $\operatorname{Cov}(\varepsilon_k)=0$ and
the covariance is preserved exactly; only the mean shifts.
\end{corollary}
\begin{proof}
Expanding $g_k$ and taking expectations,
$\mathbb E[g_k]=\operatorname{tr}(S^{-1}\!\operatorname{Cov}(r_k^{\mathrm{nom}}))
+2\,\mathbb E[(r_k^{\mathrm{nom}})^\top S^{-1}\Delta r_k]+\|\Delta r_k\|_{S^{-1}}^2$.
The first term is $\operatorname{tr}(I_p)=p$; the cross term vanishes because
$r_k^{\mathrm{nom}}$ is zero-mean and, by past-measurability, independent of $\Delta r_k$.
Since $S$ is fixed by the Riccati recursion and
$\operatorname{Cov}(\varepsilon_k)=O(\|\Delta r\|^2)$ for a bounded injection,
$\operatorname{Cov}(\tilde r_k)=S+C\operatorname{Cov}(\varepsilon_k)C^\top=S+O(\|\Delta r_k\|^2)$.
\end{proof}

\begin{corollary}[Stability dichotomy and scope]\label{cor:stability_dichotomy}
Under Theorem~\ref{thm:accumulation} with constant bias $\bar\delta$, the accumulated bias is unbounded iff $A$ has a unit-circle mode excited by the bias ($c=\psi^\top K\bar\delta\ne0$): on stable modes ($|\lambda|<1$) it saturates at $-c\lambda/(1-\lambda)$ (negligible); on marginal modes ($|\lambda|=1$) it grows polynomially; on unstable modes ($|\lambda|>1$) it would grow geometrically, but holding $\Delta r_k=\bar\delta$ then needs $a_k=\bar\delta-C\varepsilon_k$ (Proposition~\ref{prop:dynamic_necessity})
with $\|\varepsilon_k\|$ geometric, so the injection leaves the nominal manifold and forfeits stealth. The estimator-error matrix stays $A(I-KC)$, $\rho<1$: divergence is of the \emph{true state} through the
open-loop marginal mode
.
\end{corollary}

\begin{remark}
\label{rem:impact_certificate}
Theorem~\ref{thm:accumulation} shows the induced bias is not reset each step: a per-sample residual statistically indistinguishable from nominal (Corollary~\ref{cor:cov_preservation}) is compatible with an arbitrarily large accumulated deviation, the amplification being a property of the plant's marginal modes rather than of the injection. Together these form the basis for the model-free design:
maximizing the observable proxy $\mathcal L_{\mathrm{impact}}=\mathbb E\{\|\tilde y_k-y_k\|^2\}$
(Section~\ref{sec:lagrangian_objective}) subject to the stealth budget drives up a certified lower bound on $\mathbb E\{\|\Delta x_k\|^2\}$, sustained over the attack horizon. The accumulation result is stated for the linear specialization, where it admits
closed form; the per-sample certificate~\eqref{eq:residual_impact_bound} of
Lemma~\ref{lem:residual_impact_coupling} requires no such restriction.
\end{remark}

The second half of the objective~\eqref{eq:abstract_problem}, $g_k\le\eta$, depends on the detector's innovation covariance $S_k$ and is not directly enforceable by a model-free attacker. We introduce a stealth-consistent anchor $\Pi$ that recasts the stealth test as a bound on the attacker's own displacement about $\Pi$, reducing stealth to a magnitude constraint on an observable quantity; the residual budget itself is then calibrated from nominal data (below).

\begin{figure}[t]
\centering
\begin{subfigure}[b]{0.63\columnwidth}
\centering
\resizebox{\linewidth}{!}{%
\begin{tikzpicture}[scale=0.6]
\def\Mpts{(0.00,0.15) (1.00,0.62) (2.00,0.92) (3.00,1.00) (4.00,0.86) (5.00,0.52) (6.00,0.10)}
\def\Mup{(0.00,0.60) (1.00,1.07) (2.00,1.37) (3.00,1.45) (4.00,1.31) (5.00,0.97) (6.00,0.55)}
\def\Mdn{(0.00,-0.30) (1.00,0.17) (2.00,0.47) (3.00,0.55) (4.00,0.41) (5.00,0.07) (6.00,-0.35)}
\def\Mdnrev{(6.00,-0.35) (5.00,0.07) (4.00,0.41) (3.00,0.55) (2.00,0.47) (1.00,0.17) (0.00,-0.30)}
\fill[shellcol]
  plot[smooth,tension=0.7] coordinates {\Mup} --
  plot[smooth,tension=0.7] coordinates {\Mdnrev} -- cycle;
\draw[shelledge,densely dashed,line width=0.45pt] plot[smooth,tension=0.7] coordinates {\Mup};
\draw[shelledge,densely dashed,line width=0.45pt] plot[smooth,tension=0.7] coordinates {\Mdn};
\draw[Mcol,line width=1.0pt] plot[smooth,tension=0.7] coordinates {\Mpts};
\node[lbl,Mcol] at (6.38,0.08) {$\mathcal{M}$};
\coordinate (anch)  at (2.00,0.92);
\coordinate (ytrue) at (2.07,1.26);
\coordinate (yatt)  at (4.10,1.04);
\node[pt] at (anch) {};
\node[opt] at (ytrue) {};
\node[pt,fill=dcol] at (yatt) {};
\node[lbl] at (1.00,0.55){$\Pi(y_k)$};
\node[lbl,above left=-1pt and -2pt of ytrue] {$y^a_k$};
\node[lbl,dcol,above right=-1pt and 0pt of yatt] {$\tilde y^a_k$};
\draw[vec,black!70] (anch) -- (ytrue) node[lbl,pos=0.5,left=1pt] {$\delta_k$};
\draw[vec,dcol]     (anch) -- (yatt)  node[lbl,pos=0.55,below=2pt] {$d_k$};
\draw[vec,acol,densely dashed] (ytrue) -- (yatt)
      node[lbl,acol,pos=0.62,above=6pt] {$a_k=d_k-\delta_k$};
\draw[decorate,decoration={brace,amplitude=3pt,mirror},line width=0.4pt]
  (5.52,0.315) -- (5.52,0.765);
\node[lbl] at (5.92,0.54) {$\tau$};
\end{tikzpicture}}
\caption{Anchor geometry.}
\label{fig:anchor_geometry}
\end{subfigure}
\hfill
\begin{subfigure}[b]{0.35\columnwidth}
\centering
\resizebox{\linewidth}{!}{%
\begin{tikzpicture}[scale=0.4]
\coordinate (c) at (0,0);
\def\rot{22}
\def\aa{2.10}
\def\bb{1.05}
\begin{scope}[rotate=\rot]
  \fill[shellcol] (0,0) ellipse [x radius=\aa, y radius=\bb];
  \fill[okcol!14]  (0,0) circle [radius=\bb];
  \draw[line width=0.9pt] (0,0) ellipse [x radius=\aa, y radius=\bb];
  \draw[okcol,densely dashed,line width=0.7pt] (0,0) circle [radius=\bb];
  \draw[thinvec] (0,0) -- (0,\bb) node[lbl,pos=0.85,left=1pt] {$\rho_k$};
\end{scope}
\node[pt] at (c) {};
\node[lbl,below left=-5pt and -18pt of c] {$\Pi(y_k)$};
\node[lbl] at (2.50,1.50) {$g_k=\eta$};
\draw[line width=0.35pt] (2.02,1.16) -- (1.8,0.86);
\end{tikzpicture}}
\caption{Certified stealth budget that guarantees $g_k\le\eta$.}
\label{fig:eta_ellipse}
\end{subfigure}
\caption{Injected attack $a_k$ about the anchor $\Pi(y_k)$
within the certified stealth budget.}
\label{fig:attack_geometry}
\end{figure}
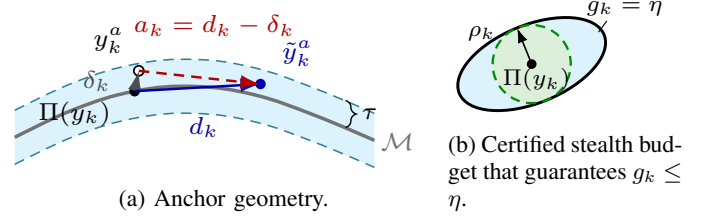

\begin{proposition}[Stealth-consistent anchor budget (R1)]
\label{prop:reflection_direction}
Let $\Pi:\mathbb R^p\to\mathbb R^p$ be a \emph{stealth-consistent anchor operator}: its output $\hat y_k:=\Pi(y_k)$ is the reference against which the residual $r_k\approx y_k-\hat y_k$ is assessed by the detector statistic $g_k=r_k^\top S_k^{-1}r_k$ (Section~\ref{subsec:system_model}), with 
$S_k\succ0$ and smallest eigenvalue $\lambda_{\min}(S_k)$. Let $\delta_k:=y_k-\hat y_k$ be the offset and $d_k$ the displacement the attack adds about the anchor, giving $\tilde y_k=\hat y_k+d_k$ and injected attack $a_k=d_k-\delta_k$ (Fig.~\ref{fig:anchor_geometry}). Any displacement within the budget
\begin{equation}
\|d_k\|\le\rho_k:=\sqrt{\eta\,\lambda_{\min}(S_k)}
\label{eq:budget}
\end{equation}
guarantees $g_k\le\eta$ for every direction of $d_k$; equivalently, $\rho_k$ is the largest isotropic displacement that is stealthy in all directions (Fig.~\ref{fig:eta_ellipse}).
\end{proposition}

\begin{proof}
By the Rayleigh quotient, $r^\top S_k^{-1}r\le\|r\|^2/\lambda_{\min}(S_k)$ for all $r\in\mathbb R^p$. Since $r_k\approx\tilde y_k-\hat y_k=d_k$ (the residual is, to leading order, the displacement itself), $\|d_k\|\le\rho_k$ gives $g_k\le\rho_k^2/\lambda_{\min}(S_k)=\eta$, independent of the direction of $d_k$, with equality when $d_k$ aligns with the eigenvector of $S_k$ associated with $\lambda_{\min}(S_k)$. Displacements outside the ball but still within the level set $g_k=\eta$ remain stealthy but are not certified by this norm bound.
\end{proof}

\begin{remark}
The budget~\eqref{eq:budget} holds for \emph{any} anchor $\Pi$ and \emph{any} (linear or nonlinear) measurement map $h$, since the argument uses only the Euclidean geometry around $\hat y_k$. At the symmetric operating point $\rho_k=\|\delta_k\|$, the budget boundary coincides with the residual-reflection construction of~\cite[Thm.~3]{guo2022stealthy}, here generalized from their linear-Gaussian, model-known setting to arbitrary nonlinear $h$. 
\end{remark}

\emph{Observable surrogate.} Since $\rho_k$ still depends on $S_k$ which require the plant model, the attacker cannot evaluate it directly. We instead enforce the budget through the observable anchor residual $r_k^{\mathrm{att}}:=\tilde y_k^a-\Pi(\tilde y_k)$, compared against a threshold $\tau$ calibrated as the $(1-\epsilon)$ percentile of $|y_k^a-\Pi(y_k)|$ on nominal data, so that $\Pr(|r_k^{\mathrm{att}}|>\tau)\le\epsilon$ under nominal
statistics. Both~\eqref{eq:budget} and its surrogate are thus fixed at the same false-alarm operating point $\epsilon$, the surrogate replacing a model-dependent covariance by an empirical quantile of the same nominal statistic.

\subsubsection{Autoencoder as a stealth-consistent manifold anchor}
\label{sec:ae_anchor}

An AE trained on nominal measurements provides a practical data-driven surrogate for the measurement manifold $\mathcal{M}$ of Lemma~\ref{lem:static}. In the linear setting, the reconstruction map of a linear AE implements orthogonal projection onto a principal subspace of the data, coinciding with Principal Component Analysis
under standard conditions~\cite{baldi1989pca}. When the nominal measurements follow the linear model $y=Hx+v$, that principal subspace is exactly
$\mathrm{col}(H)=\mathcal{M}$ (Lemma~\ref{lem:static}), so
$\mathrm{AE}(y)$ recovers $\Pi(y)$ and the reconstruction error is the Euclidean distance to $\mathcal{M}$.

In nonlinear settings, regularized (contractive or denoising) autoencoders capture the local structure of the data-generating distribution. For a reconstruction $z(y)=g(f(y))$ trained to minimize the regularized loss
\begin{equation}
\mathcal L(z)=\mathbb E\big[\,\|z(y)-y\|^2
+\sigma^2\,\|\partial z(y)/\partial y\|_F^2\,\big],
\label{eq:rcae_loss}
\end{equation}
with penalty scale $\sigma^2>0$, the optimal reconstruction satisfies, as
$\sigma^2\to0$,
\begin{equation}
z(y)-y=\sigma^2\,\nabla_y\log p(y)+o(\sigma^2),
\label{eq:score}
\end{equation}
i.e.\ the reconstruction residual estimates the \emph{score}
$\nabla_y\log p$ of the nominal measurement density
$p$~\cite{alain2014regularized,vincent2011score}. The residual therefore points
toward higher-density regions and vanishes to first order on $\mathcal{M}$
itself, where $p$ is ridged. Equivalently, the reconstruction term forces
$z(y)=y$ along directions of data variation while the penalty
in~\eqref{eq:rcae_loss} suppresses sensitivity in the orthogonal directions,
driving the singular values of $\partial z/\partial y$ toward one along the
tangent of $\mathcal{M}$ and toward zero normal to
it~\cite{vincent2011score}. Since $p$ concentrates on $\mathcal{M}$ under attack-free operation,~\eqref{eq:score} yields the manifold-distance proxy
\begin{equation}
\|y-\mathrm{AE}^\star(y)\|^2\approx\mathrm{dist}(y,\mathcal M)^2 ,
\label{eq:ae_score}
\end{equation}
which is precisely the anchor offset $\delta_k=y_k^a-\Pi(y_k)$ of
Fig.~\ref{fig:anchor_geometry}: nominal measurements lie on $\mathcal{M}$
($\delta_k\approx0$), while an off-manifold measurement is pulled back onto
$\mathcal{M}$ by the reconstruction, its residual pointing along the manifold
normal.

Property~\eqref{eq:ae_score} is exactly what
Proposition~\ref{prop:reflection_direction} requires of the anchor $\Pi$: a map
whose output serves as the reference point against which the residual is
assessed, and whose fixed set is the nominal-consistent measurement manifold.

\begin{lemma}[The autoencoder is a stealth-consistent anchor]
\label{lem:ae_anchor}
Let $\mathrm{AE}^\star$ minimize the regularized reconstruction
loss~\eqref{eq:rcae_loss} over nominal measurements, and set
$\Pi:=\mathrm{AE}^\star$. Then, to leading order in the regularization scale
$\sigma^2$, $\Pi$ is a stealth-consistent anchor in the sense of
Proposition~\ref{prop:reflection_direction}: (i) its fixed set
$\{y:\Pi(y)=y\}$ coincides with $\mathcal{M}$ up to $o(\sigma^2)$
by~\eqref{eq:ae_score}, so nominal measurements produce near-zero residuals;
and (ii) the induced residual $\delta:=y-\Pi(y)$ estimates the direction and
magnitude of the displacement from $\mathcal{M}$
\end{lemma}

\subsubsection{AE Bottleneck dimension and the observability limit}
\label{sec:bottleneck}

Let $n_z$ be the dimension of any latent bottleneck used in $AE$ or $G_\theta$.
Its correct value is set not by the state dimension (be $n_x$), but by how much of the state $x_k$ is observable from the sensor set available to the anchor,
$(y^a,\xi)$. A sufficient condition for local weak observability at $x^0$ is the
Hermann--Krener rank condition~\cite{hermann1977nonlinear}: the observability
co-distribution $d\mathcal G(x)$, spanned by the differentials of iterated Lie
derivatives of $(h_a,h_\xi)$ along admissible vector fields, has full rank $n_x$
at $x^0$; the converse holds generically~\cite[Thm.~3.11]{hermann1977nonlinear}.
When instead $\operatorname{rank} d\mathcal G(x)=q<n_x$, the missing $n_x-q$
directions leave no signature in the available measurement channel and cannot be
recovered by any encoder, which results in an observability limit. Writing $n_{\mathcal M}$ for the intrinsic dimension of the measurement manifold reachable from $(y^a,\xi)$, which coincides with $q$ when $h$ is an immersion on
the observable subspace, the design target for the bottleneck is therefore
\begin{equation}
n_z = n_{\mathcal M}
    = \operatorname{rank} d\mathcal G(x) \;\le\; n_x .
\label{eq:bottleneck_target}
\end{equation}
Setting $n_z>n_{\mathcal M}$ recovers no additional state information. Setting
$n_z<n_{\mathcal M}$ discards observable structure the anchor could otherwise
exploit. The gap $n_x-n_{\mathcal M}$ is therefore not a deficiency of the autoencoder but a measure of how much of the state is structurally invisible to the sensor set.

Since $n_{\mathcal M}$ is unavailable in closed form, we estimate it model-free via the maximum-likelihood intrinsic-dimension estimator of~\cite{levina2004maximum} on nominal $(y^a,\xi)$ windows and confirm it with
a reconstruction-error plateau over $n_z$~\cite{lee2020model,botteghi2026deep}; both fix the bottleneck at
the observable dimension.

\subsection{Structure of the Attack Law}
\label{sec:attack_law}

Requirements R1 and R2 fix \emph{what} the generator must optimize. R3 and R4 fix its \emph{structure}.

\begin{proposition}[Insufficiency of memoryless attack laws (R3)]
\label{prop:dynamic_necessity}
Consider the closed loop of Section~\ref{subsec:system_model} under an injection sequence $a_0,a_1,\dots$. An \emph{attack law} generates $a_k$ from information available at time $k$; it is \emph{memoryless} if $a_k=\phi(y_k)$ depends only on the current measurement, and \emph{causal recursive} if it maintains an internal state
\begin{equation}
s_k = G_\theta(s_{k-1},y_k), \qquad a_k = W_\theta(s_k).
\label{eq:recursive_law}
\end{equation}
Holding the residual deviation at a constant $\Delta r_k=\bar\delta$ while the impact accumulates (Theorem~\ref{thm:accumulation}) requires, from $\Delta r_k=C\varepsilon_k+a_k$, the injection $a_k=\bar\delta-C\varepsilon_k$---a functional of the \emph{accumulated} deviation $\varepsilon_k$.
\end{proposition}

\begin{proof}
 In the exact-stealth limit $\bar\delta\to0$: the admissible state offset $c_k$ (the shift the injection imprints on $\hat x$) must evolve along the free plant dynamics,
\begin{equation}
c_{k+1}=A\,c_k,
\label{eq:invariant_zero}
\end{equation}
so that the shifted trajectory remains a valid plant trajectory and leaves the recursive residual unchanged~\cite{pasqualetti2013attack, teixeira2015secure}. A \emph{constant} offset is thus dynamically stealthy only if $Ac=c$, i.e.\ $c\in\ker(A-I)$---the unit-eigenvalue (integrator) subspace; in every other direction the offset must itself evolve by~\eqref{eq:invariant_zero}. A memoryless law $a_k=\phi(y_k)$ computes the injection from the current measurement alone---a noise-corrupted projection of the current state through $h$---and cannot represent the dependence on the accumulated $\varepsilon_k$ (equivalently, cannot realize~\eqref{eq:invariant_zero}); the compensation a model-aware adversary would apply, $a_{k+1}=-CA\,\Delta e_k$, is precisely this history functional. The recursive law~\eqref{eq:recursive_law} accumulates the observed history into $s_k$ through the same causal structure that generates $\varepsilon_k$
. Table~\ref{tab:ablation} reports the closed-loop consequence of removing recurrence.
\end{proof}

Recurrence handles the temporal coupling. A second coupling is spatial: distinct sensor channels driven by a shared state are statistically dependent, so an attack on the target channel alone perturbs the joint residual. With the partition of
Section~\ref{subsec:attack_model} ($y_k=[y_k^a;\xi_k]$), let $r_k^a$ and $r_k^\xi$ denote the nominal innovation blocks for the target and context channels, with joint innovation covariance
\begin{equation}
S_k = \begin{bmatrix} S_k^a & M_k \\ M_k^\top & S_k^\xi \end{bmatrix},
\qquad M_k := \mathbb{E}\{r_k^a (r_k^\xi)^\top\}.
\label{eq:joint_cov}
\end{equation}

\begin{fact}[Cross-channel residual coupling]
\label{fact:coupling}
When multiple sensor channels are driven by a shared state $x_k$ through a joint estimator, their residuals are statistically dependent even though the measurement noises are independent: whenever the target and context channels are jointly
informative about $x_k$, the cross-covariance in~\eqref{eq:joint_cov} satisfies $M_k \ne 0$; see~\cite[Rem.~1]{guo2022stealthy} for the linear-Gaussian case, where $M_k = C^a P (C^\xi)^\top$.
\end{fact}

\begin{proposition}[Necessity of context-conditioned stealthy attacks (R4)] \label{prop:context_necessity}
Let $a_k\in\mathbb R^{p_a}$ be an attack on the target block, and suppose stealth is assessed over the full sensor set with $M_k\ne0$ (Fact~\ref{fact:coupling}). The context residual is not directly actuated---$a_k$ has no component in $\mathcal K_a^c$---but is coupled to the attack in two ways, each requiring $a_k$ to depend on $\xi_k$:
\begin{enumerate}
\item[(i)] Stealth requires the attacked joint residual to preserve the nominal cross-covariance, $\mathbb E\{\tilde r_k^a(r_k^\xi)^\top\}=M_k$ with $\tilde r_k^a=r_k^a+a_k$, equivalently $\mathbb E\{a_k(r_k^\xi)^\top\}=0$. If $a_k=\phi(y_k^a)$ uses the target channel alone, then, since $y_k^a$ and $r_k^\xi$ are
correlated through the shared state whenever $M_k\ne0$, no such $\phi$ can null this correlation for arbitrary joint statistics; the condition requires $a_k=\phi(y_k^a,\xi_k)$.
\item[(ii)] By Theorem~\ref{thm:accumulation}, the injection accumulates in $\Delta e_k$ and leaks into the context residual one step later, $\Delta r_{k+1}^\xi=(CA\,\Delta e_k)^\xi$. Any attack with impact therefore perturbs $r^\xi$ over time, so the budget must be met on the joint residual and its shaping requires the context history.
\end{enumerate}
\end{proposition}
\begin{proof}
(i) Substituting $\tilde r_k^a=r_k^a+a_k$ into the stealth requirement
$\tilde S_k=S_k$, restricted to the $(a,\xi)$ block of~\eqref{eq:joint_cov} and using $\mathbb E\{r_k^a(r_k^\xi)^\top\}=M_k$, gives $\mathbb E\{a_k(r_k^\xi)^\top\}=0$. Since $r_k^\xi$ is a fixed function of $\xi_k$ and $\hat x_{k|k-1}$ while $\mathrm{Cov}(y_k^a,r_k^\xi)\ne0$ whenever $M_k\ne0$, a map of $y_k^a$ alone cannot in general enforce this orthogonality; access to $\xi_k$ is required so that $\phi$ may be shaped---by training against the joint nominal distribution---to compensate the correlation it would otherwise induce. (ii) is immediate from Theorem~\ref{thm:accumulation} on partitioning the innovation deviation conformally with $\{\mathcal K_a,\mathcal K_a^c\}$.
\end{proof}

\paragraph*{Causal, context-conditioned anchor and generator.}\label{sec:generator_design}

We realize $\mathrm{AE}^\star$ as an anchor $\Pi$ that is a \emph{recurrent,
context-conditioned} predictor of the target channel from strictly past target
history and the currently available uncompromised context, trained on nominal data alone and held frozen thereafter.

\begin{equation}
\begin{aligned}
\hat{y}_k^a
&= \Pi(y_k), \qquad \delta_k=y_k^a-\hat{y}_k^a,\\
y_k
&=
\bigl(
y_{k-1}^a,\dots,y_{k-L}^a;\;
\xi_k,\dots,\xi_{k-L}
\bigr),
\end{aligned}
\label{eq:causal_anchor}
\end{equation}

The anchor supplies the correct \emph{recentering} of the output space onto
$\mathcal{M}$; by (R3)--(R4) the \emph{displacement} added on top of it must
itself be history- and context-dependent. We therefore parameterize the
generator as producing a displacement about the anchor of~\eqref{eq:causal_anchor}
rather than a raw measurement. The generator, also recurrent, maintains an internal state $s_k$,
is driven by a latent $z_k\sim\mathcal N(0,I)$, and emits a displacement $d_k$,
\begin{equation}
s_k = G_\theta(s_{k-1},\,y_k),
\qquad
d_k = W_\theta(s_k,z_k),
\qquad
\tilde{y}_k^a = \hat y_k^a + d_k,
\label{eq:generator}
\end{equation}
where $y_k=[y_k^a;\xi_k]$ carries both the target and the uncompromised context
blocks. The injected attack is $a_k=\tilde y_k^a-y_k^a=d_k-\delta_k$. The
recurrent state $s_k$ supplies the memory required by (R3); the context $\xi_k$
supplies the conditioning required by (R4); and the latent $z_k$ makes the policy
stochastic, so the attack is a distribution rather than a single deterministic
sequence. This parameterization recovers the static displacement $d_k$, so the
network need only learn the stealth-consistent \emph{correction} to it rather
than the full attack. The recentering also aids training: the AE-residual stealth
penalty (Section~\ref{sec:lagrangian_objective}) regularizes $d_k$ toward $0$,
i.e.\ toward the manifold, so the generator explores a compact
stealth-consistent neighborhood of $\mathcal{M}$ rather than the full measurement
space.

Furthermore, since by (R4) a stealthy attack must preserve the \emph{joint} nominal statistics of $(y^a,\xi)$, i.e. distributional stealth. 
The joint dependence is \emph{learned} from nominal data using an adversarial critic $D_\psi$ which measures the discrepancy between the generated attack and nominal joint distributions and drives the former toward the latter. We adopt a conditional Wasserstein critic with gradient penalty
(cWGAN-GP)~\cite{gulrajani2017wgangp}, conditioned on the context $\xi_k$, whose objective approximates the Wasserstein-1 distance between the generated and nominal conditional distributions. 
\begin{equation}
\mathcal{L}_{\mathrm{adv}}
= \min_{\theta}\max_{\psi\in\mathcal{D}}\;
\mathbb{E}_{(y_k^a,\xi_k)}\!\left[D_\psi(y_k,\xi_k)\right]
-\mathbb{E}_{\xi_k,\,z_k}\!\left[D_\psi(\tilde y_k,\xi_k)\right],
\label{eq:cond_wgan_objective}
\end{equation}
where $y_k=[y_k^a;\xi_k]$ is the nominal window, $\tilde y_k=[\tilde y_k^a;\xi_k]$ is the attacked window from~\eqref{eq:generator}, and $\mathcal{D}$ is the set of $1$-Lipschitz critics enforced via the gradient penalty. The Wasserstein form is preferred over a classifier discriminator~\cite{goodfellow2020generative} because it provides usable gradients even when the two distributions have limited overlap, which is the operative regime here, since the generator is confined to a narrow stealth-consistent neighborhood of $\mathcal{M}$. 


\subsection{Assembled Training Objective}
\label{sec:lagrangian_objective}

Lemma~\ref{lem:residual_impact_coupling} and
Propositions~\ref{prop:reflection_direction},~\ref{prop:dynamic_necessity},~\ref{prop:context_necessity}
each specify a requirement and its observable realization; the training objective is the weighted sum of the corresponding terms.

\emph{Impact.} By Lemma~\ref{lem:residual_impact_coupling} (R2), the attacker optimizes the observable proxy
\begin{equation}
\mathcal L_{\mathrm{impact}}
:=\mathbb E\big\{\|\tilde y_k^a - y_k^a\|^2\big\}
=\mathbb E\big\{\|d_k-\delta_k\|^2\big\},
\label{eq:impact_proxy}
\end{equation}
which induces the certified lower bound~\eqref{eq:residual_impact_bound} on the true, unobservable impact $\mathbb E\{\|\Delta x_k\|^2\}$.

\emph{Stealth.} By Proposition~\ref{prop:reflection_direction} (R1), stealth is the observable budget on the anchor residual, enforced as the one-sided penalty on the violation $h_k:=|r_k^{\mathrm{att}}|-\tau$,
\begin{equation}
\mathcal L_{\mathrm{stealth}}:=\mathbb E\big\{(h_k)_+^2\big\}.
\label{eq:stealth_penalty}
\end{equation}

\emph{Distributional stealth.} By Proposition~\ref{prop:context_necessity} (R4), the pointwise budget alone does not preserve the joint $(y^a,\xi)$ statistics; the conditional Wasserstein critic term $\mathcal L_{\mathrm{adv}}$
\eqref{eq:cond_wgan_objective} ensure distributional stealth. Finally, the recurrent, context-conditioned generator~\eqref{eq:generator} is trained to maximize impact within the stealth budget by minimizing
\begin{equation}
\min_{\theta}\;
{\mathcal L_{\mathrm{adv}}(\theta)}
\;+\;
\lambda_{\mathrm{s}}\,
{\mathcal L_{\mathrm{stealth}}}
\;-\;
{\mathcal L_{\mathrm{impact}}},
\label{eq:lagrangian}
\end{equation}
with $\lambda_{\mathrm s}$ weighting the stealth budget against impact.
Algorithm~\ref{alg:gen_fdia} details the training and deployment of the attack generator.

\begin{algorithm}[h]
\caption{Model-Free Stealthy Attack Generator}
\label{alg:gen_fdia}
\footnotesize
\begin{algorithmic}[1]
\Statex \textit{Offline training} on recorded nominal data $\{y_k\}_{k=1}^{N}$, $y_k=[y_k^a;\xi_k]$
\State $\Pi\gets\arg\min\mathbb E\|y_k^a-\Pi(y_k)\|_2^2$; freeze $\Pi=\mathrm{AE}^\star$ \Comment{\eqref{eq:causal_anchor}}
\State $\tau\gets(1-\epsilon)$ percentile of $|y_k^a-\Pi(y_k)|$ on nominal data \Comment{\eqref{eq:stealth_penalty}}
\State Init $G_\theta$ (state $s_k$, head $W_\theta$), critic $D_\psi$
\For{each iteration}
    \State $\hat y_k^a\gets\Pi(y_k)$, $\;\delta_k\gets y_k^a-\hat y_k^a$
    \State $s_k\gets G_\theta(s_{k-1},y_k,z_k)$, $\;d_k\gets W_\theta(s_k)$, $\;\tilde y_k^a\gets\hat y_k^a+d_k$, $\;\tilde y_k\gets[\tilde y_k^a;\xi_k]$
    \State $\mathcal L_{D}\gets\mathbb E[D_\psi(y_k,\xi_k)]-\mathbb E[D_\psi(\tilde y_k,\xi_k)]$; update $D_\psi$ \Comment{\eqref{eq:cond_wgan_objective}}
    \State $\mathcal L_{\mathrm{adv}}:=-\mathbb E[D_\psi(\tilde y_k,\xi_k)]$ \Comment{\eqref{eq:cond_wgan_objective}}
    \State $r_k^{\mathrm{att}}\gets\tilde y_k^a-\Pi(\tilde y_k)$; $\;\mathcal L_{\mathrm{stealth}}\gets\mathbb E[(|r_k^{\mathrm{att}}|-\tau)_+^2]$ \Comment{\eqref{eq:stealth_penalty}}
    \State $\mathcal L_{\mathrm{impact}}\gets\mathbb E\|d_k-\delta_k\|_2^2$ \Comment{\eqref{eq:impact_proxy}}
    \State $\mathcal L_G\gets\mathcal L_{\mathrm{adv}}+\lambda_{\mathrm{s}}\mathcal L_{\mathrm{stealth}}-\mathcal L_{\mathrm{impact}}$; update $G_\theta$ \Comment{\eqref{eq:lagrangian}}
\EndFor
\vspace{0.2em}
\Statex \textit{Online deployment} of trained attack model, deployed at ($k\ge k_0$)
\State Init $s_{k_0-1}$
\For{each $k$}
    \State $\hat y_k^a\gets\Pi(y_k)$, $\;s_k\gets G_\theta(s_{k-1},y_k,z_k)$, $\;d_k\gets W_\theta(s_k)$, $\;\tilde y_k^a\gets\hat y_k^a+d_k$
    \textbf{inject:} $\tilde y_k^a$ on the compromised channel.
\EndFor
\end{algorithmic}
\end{algorithm}





\section{Experimental Validation}\label{sec:experiments}

We validate the framework across three settings of increasing realism: a Model-in-the-Loop (MiL) benchmark with known dynamics (ground truth), an offline evaluation on real heavy-vehicle CAN data (transfer beyond
simulation), and a real-time Hardware-in-the-Loop (HiL) deployment (feasibility). Settings one and three exercise the full offline-train/online-deploy pipeline (Algorithm~\ref{alg:gen_fdia}); the second is offline only. Together they show the \emph{model-free} generator is (i) hazardous in closed loop, (ii) evasive against deployed IDSs, and (iii) realizable on real-time hardware.

\subsection{Model-in-the-loop validation}\label{sec:setup}
A benchmark IoT electric-vehicle CPS~\cite{rana2019iot} with fully known state-space model is selected as a baseline to validate each theoretical claim against a known reference. We use it to verify the AE based anchor (Sec.~\ref{sec:ladder}) and the certified residual--impact floor (Sec.~\ref{sec:prop2eval}), demonstrate a stealthy safety-critical hazard, with use case of lane departure, in closed loop deployment (Sec.~\ref{sec:deploy}), along with ablation study for each objective term and architecture variation of the framework (Sec.~\ref{sec:ablation}), and evaluating attack evasion against residual-based detectors as well as five state-of-the-art IDSs (Sec.~\ref{sec:ids_eval}).

The benchmark model is discretized at $T_s=1$~ms. 
The state $x_k=[\beta_k,\ \dot{\psi}_k,\ \psi_k,\ y_k^{\mathrm{lat}}]^\top$ comprises the sideslip angle $\beta$, yaw rate $\dot\psi$, yaw angle $\psi$, and lateral offset $y^{\mathrm{lat}}$; the measurement $y_k=[\dot\psi_k,\ \psi_k,\ y_k^{\mathrm{lat}}]^\top$
comprises yaw rate $\dot\psi_k$, yaw angle (heading) $\psi_k$ and lateral offset $y_k^{\mathrm{lat}}$
. Model parameters follow Table~\ref{tab:sim_params_veh}.
A discrete-time LQR regulator with feedforward gain tracks a commanded lateral maneuver, and an extended Kalman filter (EKF) provides state estimates and the innovation sequence used by the residual-based detector. The attacker compromises the \emph{lateral-offset} sensor $y^{\mathrm{lat}}$ to target lateral control of the vehicle. All the following results therefore target $y^{\mathrm{lat}}$, and the generator is trained throughout with the training objective~\eqref{eq:lagrangian}.

\begin{table}[t]
\centering
\caption{Benchmark vehicle CPS parameters~\cite{rana2019iot}.}
\label{tab:sim_params_veh}
\begin{tabular}{llll}
\toprule
Item & Value & Item & Value \\
\midrule
$T_s$   & $1$~ms          & $M$        & $380$~kg \\
$l_r$   & $0.6$~m         & $l_f$      & $0.8$~m \\
$C_r$   & $6000$~N/rad    & $C_f$      & $6000$~N/rad \\
$V_x$   & $5.5556$~m/s    & $l_{\mathrm{pre}}$ & $5.135$~m \\
$J$     & $136.08$~N$\cdot$m/(rad/s$^2$) & & \\
\bottomrule
\end{tabular}
\end{table}

\subsubsection{Validating AE-Anchor design}\label{sec:ladder}
We verify that the effective observable dimension and the autoencoder (AE) anchor 
are faithful as per lemma~\ref{lem:ae_anchor}.

\begin{figure}[t]
    \centering

    \begin{subfigure}[t]{0.44\columnwidth}
        \centering
        \includegraphics[width=\linewidth]{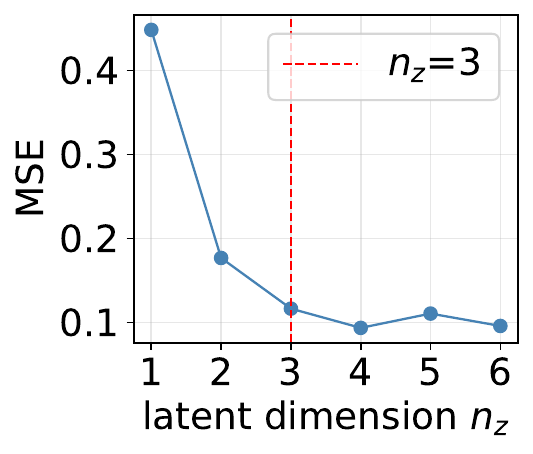}
        \caption{AE diagnostic bottleneck sweep for latent dimensions.}
        \label{fig:plateau}
    \end{subfigure}
    \hfill
    \begin{subfigure}[t]{0.54\columnwidth}
        \centering
        \includegraphics[width=\linewidth]{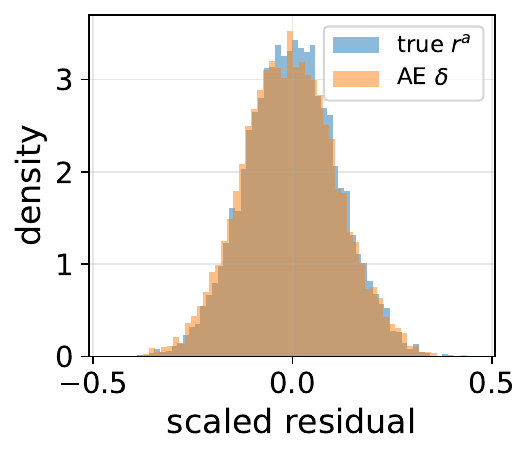}
        \caption{AE vs EKF residual variance.}
        \label{fig:ae_resid}
    \end{subfigure}
\end{figure}

\paragraph{Effective observable dimension.}
As per section~\ref{sec:bottleneck} we estimate the intrinsic dimension of the nominal measurement stream by maximum-likelihood estimation (MLE) on the raw data, with no model in the loop. The estimate converges to $3$, matching the number of independent measured channels. A diagnostic bottleneck sweep, training the AE anchor at a range of latent
dimensions, i.e. $n_z=1,\dots,6$, shows the reconstruction error collapsing from $n_z{=}1$ to $n_z{=}3$ to a plateau, shown in Fig.~\ref{fig:plateau}). Both diagnostics fix the effective observable dimension at $3$, and we set the AE bottleneck accordingly.

\paragraph{AE anchor fidelity.}
Table~\ref{tab:ae_fidelity} contrasts a memoryless reconstruction against the recurrent context-AE: the memoryless variant is nearly uncorrelated with the true innovation ($\mathrm{corr}(\delta,r_{\mathrm{EKF}})=0.25$, $\mathrm{Var}(\delta)=6.05$), whereas the recurrent context-AE recovers it almost exactly ($\mathrm{corr}(\delta,r_{\mathrm{EKF}})=0.97$, $\mathrm{Var}(\delta)=0.42$; Fig.~\ref{fig:ae_resid}). The AE residual variance matches the Kalman innovation variance to within $8\%$ ($S_{\mathrm{AE}}/S_{\mathrm{Kalman}}=1.08$), confirming
that the AE is a faithful, minimally parameterized model of the nominal manifold.

\paragraph{Threshold calibration}
Fixing the AE anchor, we calibrate the stealth threshold on nominal data at a $1\%$ target false-alarm rate. Let $\sigma_{\mathrm{AE}}$ denote the standard deviation of the AE reconstruction residual on nominal data and $\tau_{\mathrm{AE}}$ the $99$th-percentile threshold; we obtain $\sigma_{\mathrm{AE}}=0.1186$ and $\tau_{\mathrm{AE}}=0.304$ (scaled units) 
and is the stealth regularizer in the $\mathcal L_{\mathrm{stealth}}$ term of the training objective~\eqref{eq:lagrangian}.

\begin{table}[t]
\centering
\caption{AE anchor fidelity against the true EKF innovation.}
\label{tab:ae_fidelity}
\small
\begin{tabular}{lcc}
\toprule
Variant & $\mathrm{Var}(\delta)$ & $\mathrm{corr}(\delta,r_{\mathrm{EKF}})$ \\
\midrule
Memoryless        & 6.049 & 0.249\\
Recurrent + context & 0.420 & 0.969\\
\bottomrule
\end{tabular}
\end{table}

\subsubsection{Validating residual--impact coupling under stealth constraint}\label{sec:prop2eval}

We test Lemma~\ref{lem:residual_impact_coupling} directly in its stealth-constrained form. Under the stealth event $\{\|r_k\|\le\tau_r\}$, the triangle inequality $\|a_k\|\le\|r_k\|+L\|\Delta x_k\|+\sqrt{\sigma_v}\,\kappa$ gives the injection-magnitude floor $\|\Delta x_k\|\ge(\|a_k\|-\tau_r-\sqrt{\sigma_v}\,\kappa)/L$. From the
benchmark and a nominal run, $L=\|C\|_2=1.0$, $\sqrt{\sigma_v}=0.05$ ($\sigma_v=\lambda_{\max}(V)=2.5\times10^{-3}$), $p=3$, and $\tau_r=0.137$ (the $99$th percentile of the nominal $\|r_k\|$). Growing $\|a_k\|$ through a slow ramp that the filter tracks keeps the innovation below $\tau_r$ while the bias accumulates, so $\|\Delta x_k\|$ rises with $\|a_k\|$. 
Every stealthy sample lies on or above the floor and $\|\Delta x_k\|$ tracks $\|a_k\|$ linearly, confirming the attacker drives the estimation error in proportion to the injection budget; the bound holds for
essentially all samples, well above the $0.67$ Chebyshev guarantee, which bounds only how often $\mathcal E_\kappa$ occurs rather than the floor itself.
\subsubsection{Closed-loop deployment and impact validation}\label{sec:deploy}
\begin{figure}[t]
    \centering
    \begin{subfigure}[t]{0.44\columnwidth}
        \centering
        \includegraphics[width=\linewidth]{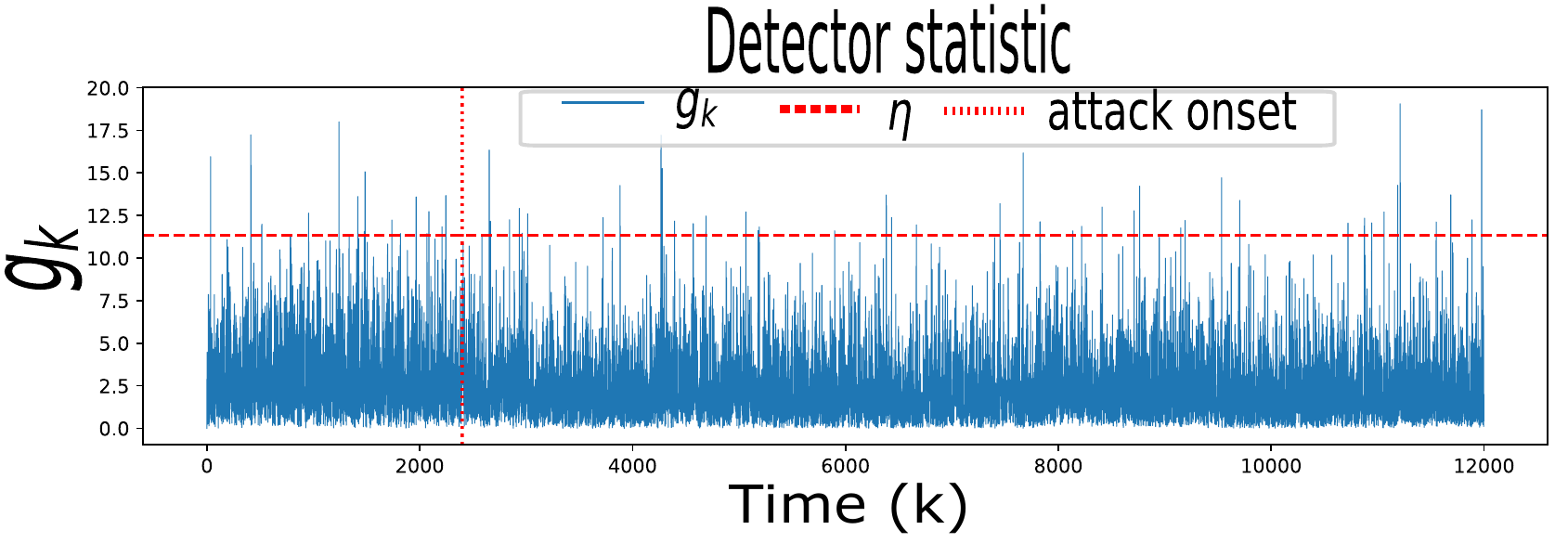}
        \caption{Detector statistic $g_k$.}
        \label{fig:gk}
    \end{subfigure}
    \hfill
    \begin{subfigure}[t]{0.54\columnwidth}
        \centering
        \includegraphics[width=\linewidth]{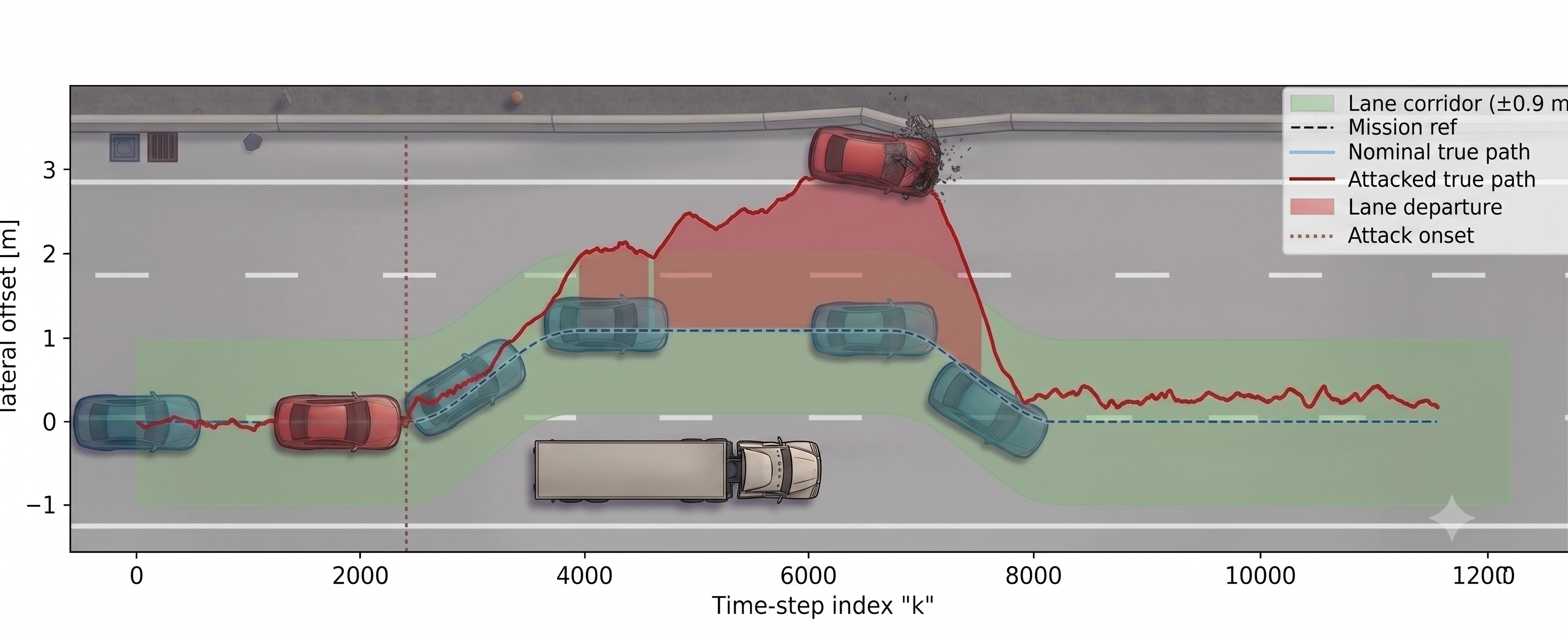}
        \caption{Visualization of attack hazard.}
        \label{fig:lane_change_attack_visual}
    \end{subfigure}
    \vspace{0.6em}
    \begin{subfigure}[t]{0.9\columnwidth}
        \centering
        \includegraphics[width=\linewidth]{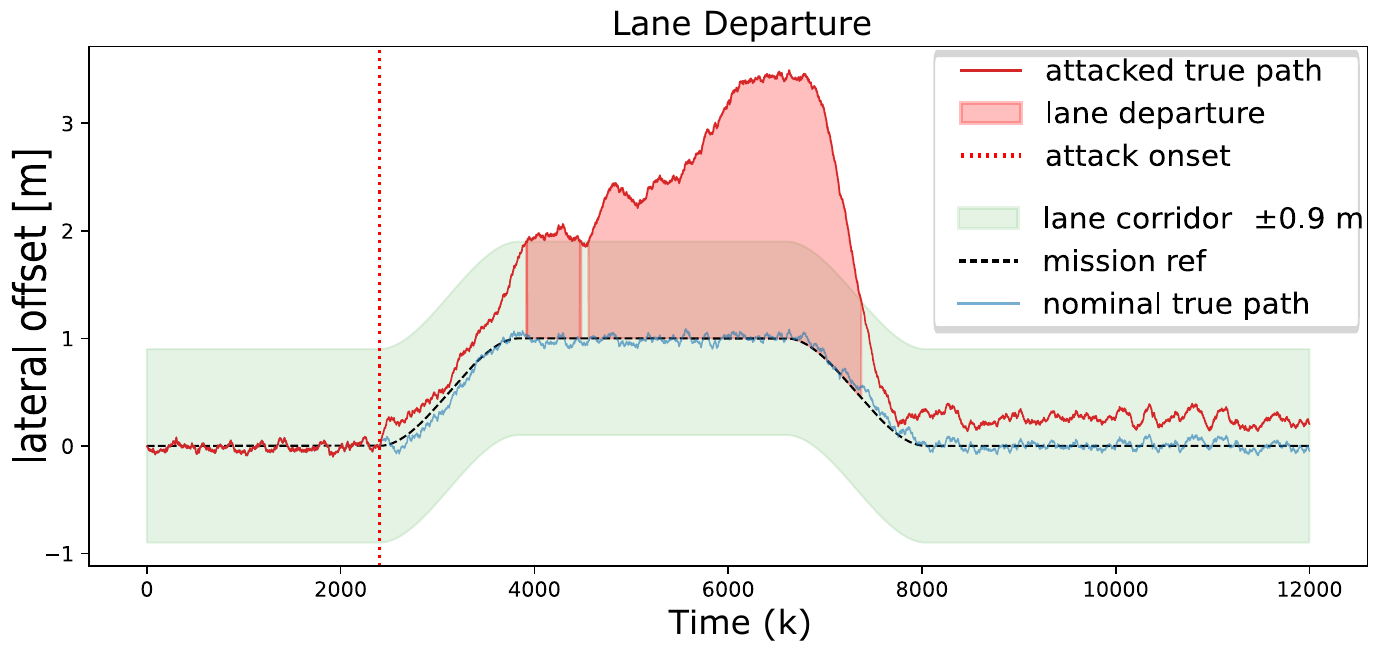}
        \caption{Vehicle lane change hazard with $2.51$~m lateral offset during attack.}
        \label{fig:departure}
    \end{subfigure}
    \caption{Vehicle lane-change mission}
    \label{fig:lane_change}
\end{figure}

We deploy the trained generator on a lane-change mission, with the attack
activated at $k_0=2400$ and all detectors calibrated to a $1\%$ false-alarm rate on the nominal run. Fig.~\ref{fig:departure} shows the true vehicle trajectory: the attacked path departs the $\pm0.9$~m lane corridor, reaching a peak lateral departure of $2.51$~m and remaining outside the corridor for $3.37$~s, while the nominal path stays centered. Crucially, the residual-based detector statistic $g_k$ remains at its nominal level throughout the attack (Fig.~\ref{fig:gk}): the $\chi^2$
alarm rate during the attack ($0.007$) is below the nominal value, and the mean statistic $\mathbb{E}[g_k]=2.21$ sits below the nominal $3.01$. The attack thus induces a safety-critical lane departure while presenting a residual signature indistinguishable from normal operation, the closed-loop realization of the coupling certified in Lemma~\ref{lem:residual_impact_coupling}.

\subsubsection{Ablation Study}\label{sec:ablation}
Table~\ref{tab:ablation} isolates the contribution of each objective term and
architectural choice, deploying every variant in closed loop on the lane-change
mission of the benchmark. Removing the stealth budget
($\mathcal{L}_{\mathrm{adv}}$, $\mathcal{L}_{\mathrm{stealth}}$, or their
combination) produces large impact but is detected outright, with alarm rates
saturating near unity. Removing the impact term
($\mathcal{L}_{\mathrm{impact}}$ alone) collapses the attack to the nominal
floor which is stealthy but without any impact, at or below the $0.205$~m nominal tracking error.
Only the full objective and its stealth-constrained pairs remain below every detector threshold while producing a lateral offset (hazardous impact): the proposed
$\mathcal{L}_{\mathrm{adv}}+\mathcal{L}_{\mathrm{impact}}+\mathcal{L}_{\mathrm{stealth}}$
achieves the largest impact ($2.51$~m peak, $0.77$~m sustained) under stealth constraint with all alarm rates at nominal. The architecture variation shows recurrence is
necessary: replacing the recurrent generator with an MLP diverges the magnitude of impact and is detected, whereas removing only the context channels
leaves the attack stealthy at reduced impact.

\begin{table}[t]
\centering
\caption{Ablation study. Nominal reference: $\chi^2$ rate $0.0091$,
$\mathbb{E}[g_k]=3.01$. S/A/I denote the $\mathcal{L}_{stealth}$,
$\mathcal{L}_{adv}$, $\mathcal{L}_{impact}$ loss terms; $^{\dagger}$ denotes
detected.}
\label{tab:ablation}
\setlength{\tabcolsep}{3pt}
\resizebox{\columnwidth}{!}{%
\begin{tabular}{lcccc}
\toprule
& \multicolumn{2}{c}{Impact} & \multicolumn{2}{c}{Detection rate} \\
\cmidrule(lr){2-3}\cmidrule(lr){4-5}
Variant & peak $\|\Delta x\|$ & avg $\|\Delta x\|$
& $\chi^2$ & $\mathbb{E}[g_k]$ \\
\midrule
S                       & 0.16    & 0.04    & 0.005 & 2.09 \\
A$^\dagger$             & 165.6   & 68.90   & 0.514 & 14.9 \\
I$^\dagger$             & 500.1   & 305.32  & 0.755 & $3.2{\times}10^{4}$ \\
A+I$^\dagger$           & 7827.3  & 4014.12 & 1.000 & $5.2{\times}10^{4}$ \\
A+S                     & 1.18    & 0.41    & 0.004 & 2.09 \\
I+S                     & 2.07    & 0.51    & 0.004 & 2.09 \\
\textbf{A+I+S}          & \textbf{2.51} & \textbf{0.77} & \textbf{0.007} & \textbf{2.21} \\
\midrule
\multicolumn{5}{l}{\emph{Architecture variants}}\\
$w/o$ context                  & 0.83 & 0.27 & 0.004 & 2.08 \\
$w/o$ recurrence (MLP)$^\dagger$ & $8.1{\times}10^{5}$ & $4.7{\times}10^{4}$ & 0.661 & $1.9{\times}10^{9}$ \\
$w/o$ context, recurrence$^\dagger$ & 4609.4 & 257.31 & 0.354 & $6.4{\times}10^{4}$ \\
\bottomrule
\end{tabular}}
\end{table}

\subsubsection{Evaluation Against SOTA IDSs}\label{sec:ids_eval}
We evaluate against five state-of-the-art IDSs---CANet~\cite{hanselmann2020canet}, INDRA~\cite{kukkala2020indra}, NeuroCAN~\cite{balaji2021neurocan}, and DL~LSTM--GNB/CNNLSTM--GNB~\cite{mansourian2023deep}---each trained on nominal data and thresholded at $1\%$ False Positive Rate
(FPR); a Gaussian-injection reference confirms all are operational, and all achieve near-optimal True Positive Rate (TPR) against known attack classes (Table~\ref{tab:ids_combined}). Against the proposed attack, CANet and INDRA, which only monitor the compromised channel, are evaded entirely ($0.1\%$), consistent with a residual-based detector leaving no exploitable signature on the compromised
channel itself. The multivariate detectors (NeuroCAN, DL~LSTM--GNB, DL~CNNLSTM--GNB) retain partial sensitivity ($\approx25\%$) through cross-channel correlation, observing the state drift the attack induces in the co-observed channels. This residual sensitivity is therefore a consequence of \emph{impact}, not a failure of stealth. Notably, the two GNB detectors are trained via supervised learning on labeled instances of this exact attack, yet still miss $\approx75\%$ of attack samples. This establishes that defending against this attack class requires a multivariate detection approach trained to identify the attack's specific
signature, which is precisely the mechanism the adversarial generator
proposed here is designed to synthesize.

\begin{table}[h]
\centering
\caption{Detection performance (TPR\,\%) of benchmark IDSs against attack classes at $1\%$ FPR. $^{\dagger}$Supervised.}
\label{tab:ids_combined}
\setlength{\tabcolsep}{3pt}
\resizebox{\columnwidth}{!}{%
\begin{tabular}{lccccccc}
\toprule
\textbf{IDS} & \textbf{Conti.} & \textbf{Spoof} & \textbf{Replay} & \textbf{Plat.} & \textbf{Supp.} & \textbf{Gauss.} & \textbf{Ours} \\
\midrule
CANet \cite{hanselmann2020canet}          & 70.1  & --    & 87.8 & 98.8 & 17.6  & 92.8 & 0.1 \\
INDRA \cite{kukkala2020indra}             & 99.0  & --    & 98.5 & 98.0 & 100.0 & 92.7 & 0.1 \\
NeuroCAN \cite{balaji2021neurocan}        & --    & 90.0  & --   & --   & --    & 93.4 & 24.4 \\
DL LSTM--GNB$^{\dagger}$\cite{mansourian2023deep}    & 100.0 & 100.0 & --   & --   & 100.0 & 98.3 & 22.7 \\
DL CNNLSTM--GNB$^{\dagger}$\cite{mansourian2023deep} & 99.7  & 100.0 & --   & --   & 100.0 & 97.8 & 27.9 \\
\bottomrule
\end{tabular}}
\end{table}

\subsection{Validation on real vehicle data}\label{sec:csu}
\begin{figure}[t]
    \centering
    \begin{subfigure}[t]{0.49\columnwidth}
        \centering
        \includegraphics[width=\linewidth]{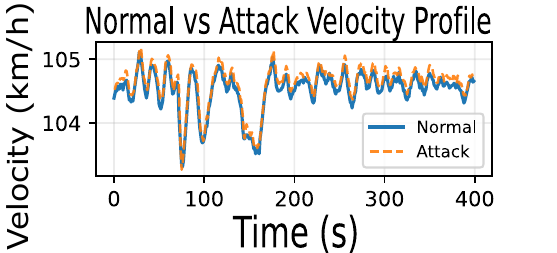}
        \caption{Normal $y$ vs.\ stealthy attack.}
        \label{fig:y_pred_vs_real}
    \end{subfigure}
    \hfill
    \begin{subfigure}[t]{0.49\columnwidth}
        \centering
        \includegraphics[width=\linewidth]{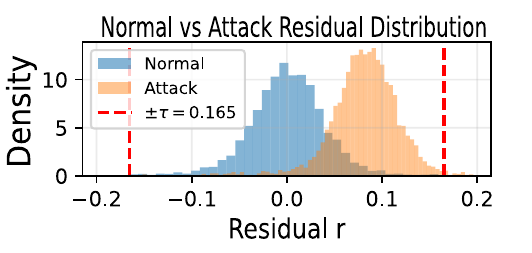}
        \caption{Residual $r$ distributions with 
        $\pm\tau_{0.99}$ threshold (dashed).}
        \label{fig:residual_dist}
    \end{subfigure}
    \caption{Stealthy attack performance on real vehicle data. 
    }
    \label{fig:csu}
\end{figure}

In the second setting, the generator is trained and evaluated offline on recorded nominal CAN measurements from a real heavy vehicle (Kenworth)~\cite{biggs2024modeling}, using held-out segments for evaluation. This tests whether the method transfers from the synthetic benchmark to real sensor data. The generator learns to generate a compromised measurement that tracks the nominal manifold closely enough to remain within the AE stealth band while carrying a controlled offset from the true value (Fig.~\ref{fig:csu}). Because the evaluation is offline with no plant or detector running in the loop, the injected offset and the anchor residual coincide. This setting establishes stealthy attack synthesis on real vehicle data and operates beyond simulation.



\subsection{Real-time hardware-in-the-loop validation}\label{sec:hil}

\begin{figure}[t]
    \centering
    \includegraphics[width=1.0\linewidth]{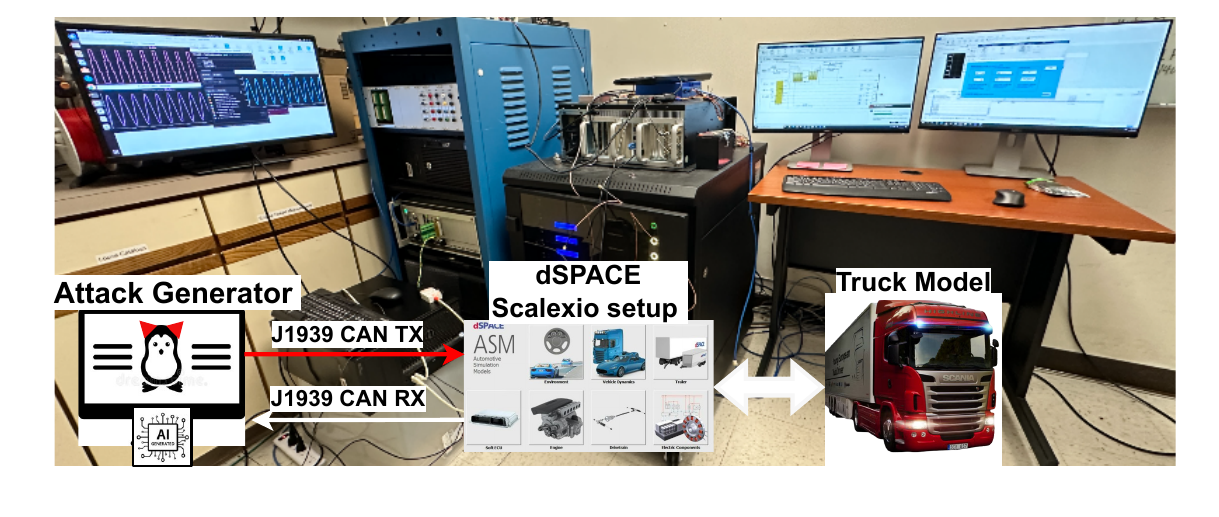}
    \caption{dSPACE HiL setup for real-time attack deployment.}
    \label{fig:dspace}
\end{figure}

Finally for validation of the attack on a real-time HiL platform, a dSPACE SCALEXIO LabBox with a DS6342 CAN board, CAN (J1939) bus (heavy-vehicle compliant and publicly available) and a high-fidelity (Scania) truck model~\cite{ERIKSSON2016121} in a closed-loop setting is used. The generator is trained on nominal data from HiL and deployed online with its forward inference pass through a Linux-based API, representative of malicious in-vehicle-network APIs used in practical CAN injection attacks \cite{Tindell2023CANInjection, jepson2024commercial}, as shown in Fig.~\ref{fig:dspace}. The HiL platform provides reproducibility and safety, so that observed deviations are attributable to the attack alone rather than to environmental or control-logic variation. Table~\ref{tab:sim_params} provides details about the road scenario and driving profile with vehicle under cruise control setting.


\begin{table}[t]
\centering
\caption{Simulation parameters and driving scenario.}
\label{tab:sim_params}
\begin{tabular}{ll}
\toprule
\textbf{Parameter} & \textbf{Value} \\
\midrule
Hill wavelength ($\lambda$)              & $2000$~m \\
Max slope ($\theta_{\max}$)              & $\pm 1.5^\circ$ \\
Velocity profile ($v_{\mathrm{ref}}$)    & $30$--$90$~km/h (staircase) \\
Step magnitude ($\Delta v$)             & $10$~km/h \\
Horizon ($T$)       & $0$--$1000$~sec \\
\bottomrule
\end{tabular}
\end{table}

\begin{figure}[t]
    \centering
    \begin{subfigure}[t]{0.54\columnwidth}
        \centering
        \includegraphics[width=\linewidth]{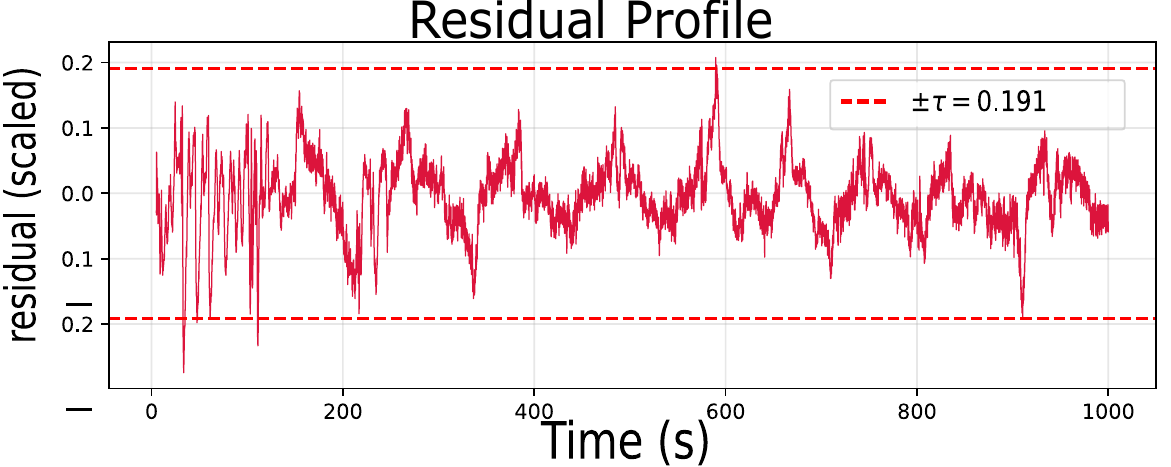}
        \caption{Detector statistic $g_k$.}
        \label{fig:hil_stealth}
    \end{subfigure}
    \hfill
    \begin{subfigure}[t]{0.44\columnwidth}
        \centering
        \includegraphics[width=\linewidth]{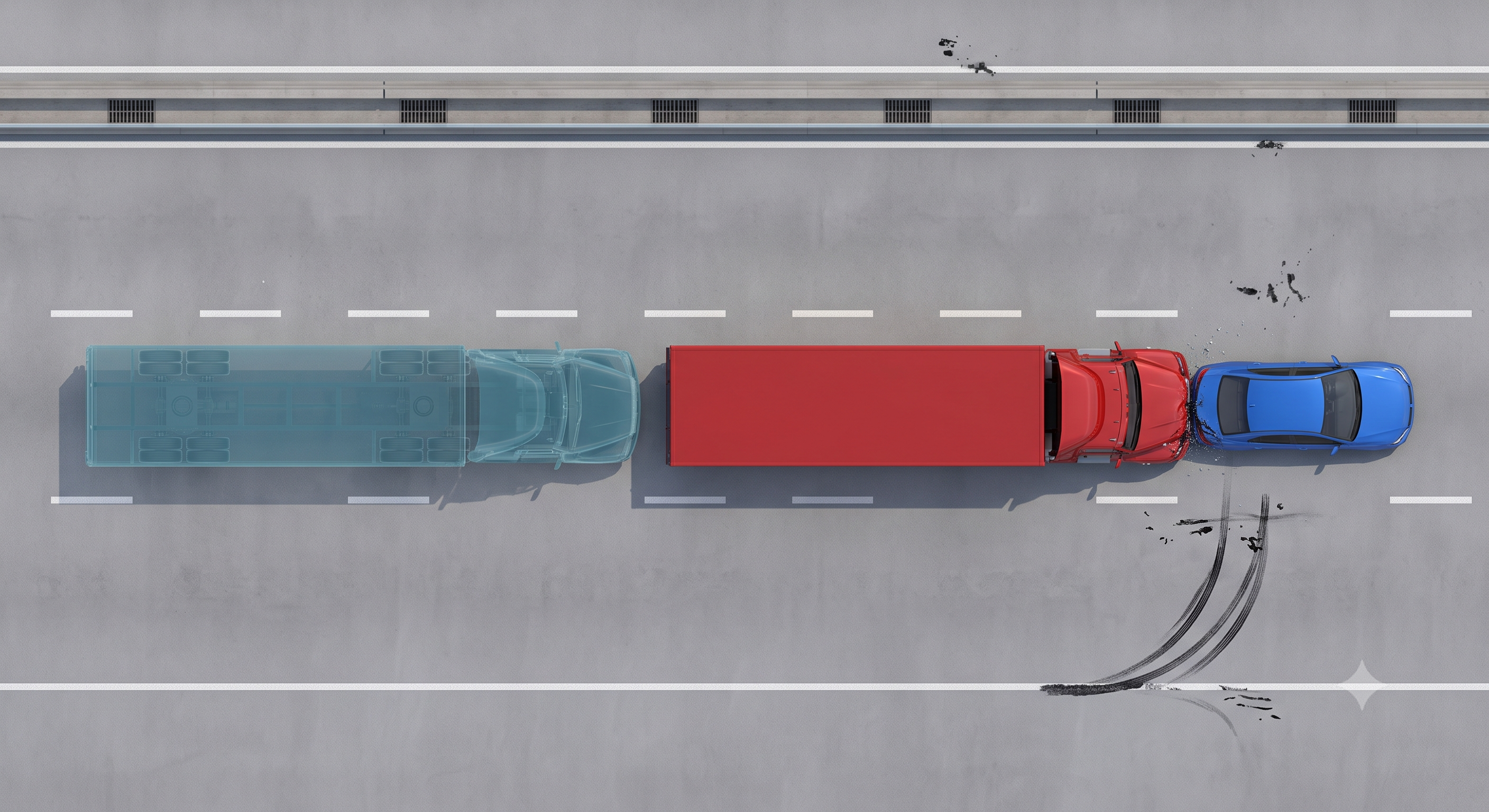}
        \caption{Visualization of attack hazard.}
        \label{fig:hil_attack_visual}
    \end{subfigure}
    \vspace{0.6em}
    \begin{subfigure}[t]{0.85\columnwidth}
        \centering
        \includegraphics[width=\linewidth]{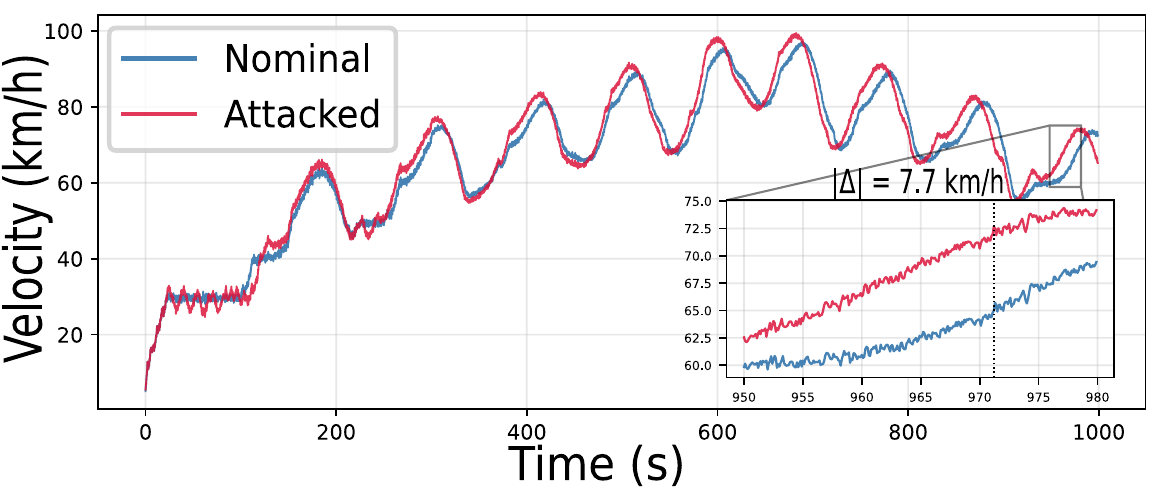}
        \caption{Vehicle speed exceeding by 7.7 km/h than nominal at 965 sec during attack.}
        \label{fig:hil_vel}
    \end{subfigure}
    \caption{Vehicle velocity profile during real-time attack.}
    \label{fig:hil}
\end{figure}

The attack targets the longitudinal dynamics of the vehicle and drives the true vehicle speed significantly off-nominal to a hazardous state while remaining below the detector threshold. The vehicle-response deviation reaches $7.7$~km/h at 965 sec with a sustained positive bias
, as shown in Fig.~\ref{fig:hil_vel}). Simultaneously, the attacked residual remains within the threshold $\tau=0.191$, with the exceedance $0.0043$, statistically indistinguishable from the nominal false-alarm rate (Fig.~\ref{fig:hil_stealth}). 
The real-time realization of the attack generator shows the feasibility of attack deployment on actual vehicle under real bus timing, quantization, and closed-loop dynamics.

\section{Conclusion}\label{sec:conclusion}

A safety-critical CPS can be driven to a hazardous state by an adversary using passively recorded nominal measurements only. Without using any plant model, estimator gain, or detector threshold, the learned generator moved a benchmark electric vehicle $2.51$\,m out of its lane and a real-time truck-in-the-loop platform $\approx\!8$~km/h off its nominal speed while keeping residual statistic at its nominal false-alarm level. This work replaces the exact algebraic invariance available only to a model-aware attacker with an observable stealth budget about a stealth-consistent anchor, which the closed loop integrates into a large physical deviation. Conclusively, defensive leverage lies in cross-channel consistency, making sensor-placement and channel-partitioning choices security decisions. This attack can further be inverted into an offense-informed, cross-channel detector, closing the loop between the threat and the defense it motivates.

\renewcommand{\bibfont}{\footnotesize}

\printbibliography

\end{document}